%% file: arXiv_main.tex
\documentclass[aps,pra,twocolumn,showpacs,preprintnumbers,amsmath,amssymb,footinbib]{revtex4-1}
\pdfoutput=1

\usepackage{float}
\usepackage{booktabs}

\input{preamble}

\begin{document}

\title{Joint Subgraph-to-Subgraph Transitions \\ Generalizing Triadic Closure for Powerful and Interpretable Graph Modeling}

\author{Justus Hibshman}
\email{jhibshma@nd.edu}
\affiliation{University of Notre Dame -- Notre Dame, Indiana 46556, USA}

\author{Daniel Gonzalez Cedre}
\thanks{Second and third authors contributed equally to this research}
\email{dgonza26@nd.edu}
\affiliation{University of Notre Dame -- Notre Dame, Indiana 46556, USA}

\author{Satyaki Sikdar}
\thanks{Second and third authors contributed equally to this research}
\email{ssikdar@nd.edu}
\affiliation{University of Notre Dame -- Notre Dame, Indiana 46556, USA}

\author{Tim Weninger}
\email{tweninge@nd.edu}
\affiliation{University of Notre Dame -- Notre Dame, Indiana 46556, USA}

\begin{abstract}
  We generalize triadic closure, along with previous generalizations of triadic closure, under an intuitive umbrella generalization: the Subgraph-to-Subgraph Transition (SST). We present algorithms and code to model graph evolution in terms of collections of these SSTs. We then use the SST framework to create link prediction models for both static and temporal, directed and undirected graphs which produce highly interpretable results. Quantitatively, our models match out-of-the-box performance of state of the art graph neural network models, thereby validating the correctness and meaningfulness of our interpretable results.
\end{abstract}

\maketitle

\input{body}
\bibliography{references}

\end{document}

%% file: preamble.tex
\usepackage{tikz,amsmath, amssymb,bm,color}
\usepackage{multirow} 
\usepackage{colortbl}
\usepackage{enumitem}
\usetikzlibrary{shapes,arrows, arrows.meta}
\usetikzlibrary{backgrounds,calc,arrows,decorations.markings, shapes,shapes.arrows,optics,patterns,hobby, fit, positioning, automata}
\usetikzlibrary{calc}
\usepackage{pgfplots}
\usetikzlibrary{pgfplots.groupplots}
\usepgfplotslibrary{dateplot,fillbetween}
\pgfplotsset{compat=newest}

\tikzstyle{textnode} = [rectangle, inner sep=0pt,outer sep=0,execute at begin node={\strut}, font=\small]  
\tikzstyle{opencircle} = [circle, thick, draw=black!80, minimum width=10, fill=black!5, inner sep=1.5, text=black]
\tikzstyle{faintopencircle} = [circle,minimum width=10, draw=black!40, fill=black!05, inner sep=1.5, text=black!40]
\tikzstyle{itxset} = [rounded corners=3pt, draw, minimum height=14pt, inner sep=0]
\tikzstyle{itxsetfocus} = [itxset, ultra thick]
\tikzstyle{internalnode} = [textnode,circle, draw=black, very thick, fill, text=black]
\tikzstyle{externalnode} = [textnode, circle, draw, thick]
\tikzstyle{nonterminal} = [textnode,draw,inner xsep=1.5]
\tikzstyle{graphletnode} = [circle, draw, fill, minimum size=1.0mm,inner sep=0pt,outer sep=0pt, fill=black]
\tikzstyle{splitnode} = [circle, draw, fill=black, minimum size=2.2mm,inner sep=0pt,outer sep=0pt]
\tikzset{ghost/.style={color=gray!50}}
\tikzstyle{faintnonterminal} = [draw=black!40, inner sep=1.5, text=black!40]
\tikzstyle{faintedge} = [very thin, color=black!50]
\tikzstyle{enode-hrg} = [circle, draw, fill=black, text=white, minimum size=4mm, inner sep=0.25pt, outer sep=1.5pt]  
\tikzstyle{hyperedge} = [draw, inner xsep=0.5, fill=gray!5, diamond, minimum size=0.5mm, outer sep=0.5pt]  

\tikzstyle{textnode} = [rectangle, inner sep=0pt,outer sep=0,execute at begin node={\strut}, font=\small]  

\tikzstyle{bnode} = [circle, draw, fill=NDblue, minimum size=2mm, text=white, outer sep=1pt]  

\tikzstyle{enode} = [circle, draw=white, thick, inner sep=1.5pt, fill=gray!20, outer sep=2pt, text=black]  

\tikzstyle{inode} = [circle, thick, minimum size=5pt, draw=black, fill=gray!10!white, inner sep=0pt, outer sep=1.5pt]  

\tikzstyle{nt} = [draw=pastelred, very thick, rectangle, inner xsep=1.5, fill=gray!5, minimum width=3mm, minimum height=5mm, outer sep=2pt, text=black, font=\large]  

\tikzstyle{tnode} = [minimum size=5mm, font=\large]  

\tikzstyle{hnode} = [enode, very thick, draw=NDgold]  

\tikzstyle{hidden} = [draw=none, rectangle]

\tikzstyle{edge} = [thick, draw=gray]  

\tikzstyle{iedge} = [edge, very thick, draw=gray]   

\tikzstyle{bedge} = [edge, draw=NDgreen]  

\tikzstyle{deledge} = [edge, draw=red, dotted]  

\tikzstyle{faded} = [opacity=0.40, text opacity=0.40]

\tikzstyle{newnode} = [circle, thick, minimum size=2mm, outer sep=1.5pt, draw=blue,fill=cyan!50]  

\tikzstyle{Arrow} = [line width=0.5mm, draw=NDgold!80, -{Triangle[length=1.55mm,width=0.95mm]}, scale=0.75, transform shape]

\tikzstyle{hedge} = [edge, ->, very thick, draw=gray!80]

\tikzstyle{nedge} = [hedge, draw=cyan!80]

\tikzstyle{redge} = [hedge, draw=red!60]

\tikzstyle{diredge} = [->, >={latex'[length=1mm]}, thick]
\tikzstyle{boundedge} = [diredge, draw=NDgold]  

\tikzset{newedge/.style = {->,> = latex', very thick}}

\tikzstyle{uedge} = [thick, shorten >=1pt, auto,]  

\tikzstyle{highlight} = [very thick, preaction={fill=gray!50}, pattern=north east lines]

\tikzstyle{shrink} = [scale=0.5, transform shape]

\tikzstyle{treenode} = [enode, shrink]

\tikzstyle{tedge} = [->, >=latex, draw=gray, thin]

\tikzstyle{tedge2} = [-{Latex[length=1mm]}, draw=gray, thin]

\tikzstyle{blob} = [ellipse, draw=black, dashed, minimum height=40pt, minimum width=37.5, outer sep=1.5pt]
\tikzstyle{dottededge} = [edge, dashed]

\tikzstyle{sourceinode} = [inode, fill=gray!80]
\tikzstyle{targetinode} = [inode, fill=black!80]

\tikzstyle{newestdiredge} = [diredge, draw=blue!50!cyan, fill=blue!50!cyan]
\tikzstyle{newdiredge} = [diredge, draw=blue!80!white, fill=blue]
\tikzstyle{olddiredge} = [diredge, draw=gray, fill=gray]
\tikzstyle{neverdiredge} = [diredge, draw=cyan, fill=cyan]

\tikzstyle{t0} = [line width=1pt]
\tikzstyle{t1} = [line width=1.5pt]
\tikzstyle{t2} = [line width=2pt]
\tikzstyle{t3} = [line width=2.5pt]

\usepackage{pgfplots}
\usetikzlibrary{pgfplots.groupplots}
\selectcolormodel{cmyk}

\newenvironment{customlegend}[1][]{%
    \begingroup
    \csname pgfplots@init@cleared@structures\endcsname
    \pgfplotsset{#1}%
}{%
    \csname pgfplots@createlegend\endcsname
    \endgroup
}%

\def\addlegendimage{\csname pgfplots@addlegendimage\endcsname}

\pgfplotsset{
    jitter/.style={
        y filter/.code={\pgfmathparse{\pgfmathresult+rnd*#1}}
    },
    jitter/.default=0.05
}
\pgfplotsset{every tick label/.append style={font=\scriptsize}}
\pgfplotsset{compat=newest}
\pgfdeclarelayer{background}
\pgfdeclarelayer{foreground}
\pgfsetlayers{background,main,foreground}   

\tikzstyle{globaloptions} = [] 
\tikzstyle{sst} = [draw=blue, fill=blue!60, globaloptions]  
\tikzstyle{sst3} = [draw=cyan, fill=cyan!60, globaloptions]  

\tikzstyle{gcnae} = [draw=red, fill=red!60, globaloptions]  
\tikzstyle{gcnvae} = [draw=red!50, fill=red!30, globaloptions]

\tikzstyle{comm} = [draw=lime, fill=lime!60, globaloptions]
\tikzstyle{random} = [draw=black, fill=black!60, globaloptions]

\tikzstyle{linearae} = [draw=green!40!black, fill=green!70!black]  
\tikzstyle{linearvae} = [draw=green!70!black, fill=green!50!white]  

\tikzstyle{gravityae} = [draw=pink, fill=pink!50!red, globaloptions]  
\tikzstyle{gravityvae} = [draw=pink!80, fill=pink!90!red, globaloptions]

%% file: body.tex
\section{Introduction}

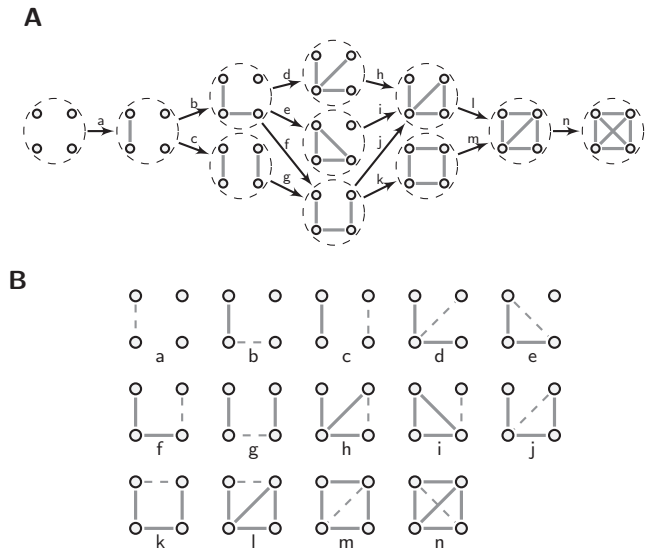
\begin{figure}
    \centering
  \input{figures/fig1}
  \caption{Edge-Addition Subgraph-to-Subgraph Transitions for Undirected Four-Node Subgraphs. The same information is depicted in two formats (\textit{A} and \textit{B}). \textit{A:} Each arrow indicates a possible subgraph-to-subgraph transition (SST) caused by adding an edge to an undirected four-node subgraph. This format helps illustrate that subgraphs may transition to (and be transitioned to from) isomorphically distinct subgraphs. \textit{B:} Each dashed missing edge indicates a possible subgraph-to-subgraph transition caused by adding the dashed edge to an undirected four-node subgraph. This format helps illustrate that each distinct edge-addition transition corresponds to an isomorphically-distinct missing edge in a subgraph.}
  \label{fig:teaser}
\end{figure}

Triadic closure is a widely known, simple process for modeling the evolution and dynamics of many real world graph processes \citep{bianconi2014triadic, klimek2013triadic}. Triadic closure's use in the graph modeling community is due, in large part, to its ability to intuitively explain commonly observed social and natural phenomenon. For example, social balance theory is built upon achieving consistency among individuals in social network triads~\citep{granovetter1977strength}, and social networks commonly predict friendship links that close the most triangles~\citep{adamic2003friends}. In addition to triads, analysis of the evolution and dynamics of other small subgraphs ({\em i.e.}, graphlets, motifs, etc.) have proven to be illuminating and pleasantly interpretable for many graph mining and scientific tasks~\citep{paranjape2017motifs, ugander2013subgraph, prvzulj2007biological, liu2020temporal}.

To this end, researchers have generalized the concept of triadic closure in different ways. For instance, Seshadhri et. al. considered the many different kinds of triangle closures possible in a directed graph~\citep{seshadhri2017directed}. Yin et. al. considered something similar to Seshadhri et. al., but did not include bidirected edges in their enumeration of directed triadic closure types~\citep{yin2019measuring}. Rossi et. al. considered ``motif closures,'' whereby they mean any occurrence of a motif being formed by the adding of an edge~\citep{rossi2020closing}.

We offer an elegant generalization which encapsulates and expands upon previous generalizations of triadic closure: The Subgraph-to-Subgraph Transition (SST). In our formulation triangle closure can be considered one specific kind of subgraph-to-subgraph transition: open-wedge to triangle. SSTs are also a generalization of ``motif closures,'' as a motif closure only considers the resulting subgraph, not the beginning subgraph (see~\citep{rossi2020closing}); thus a single motif closure may correspond to many distinct SSTs.

For example, Fig.~\ref{fig:teaser} depicts all of the possible four-node subgraph-to-subgraph transitions in two formats: (A) as state transitions, and (B) with added edges. Although not shown in Fig.~\ref{fig:teaser}, our SST algorithms can handle significantly larger subgraphs (albeit with runtime implications), over directed or undirected graphs, for both node and edge additions and deletions.

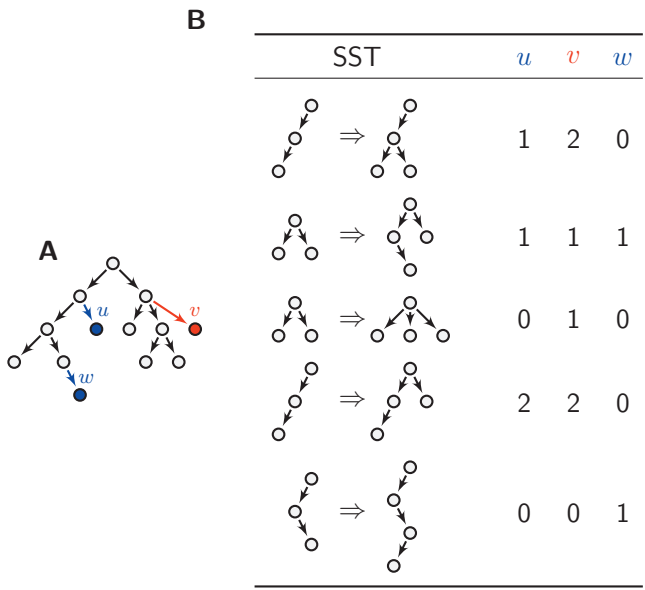
\begin{figure}
  \input{figures/fig2}
  \caption{Growing a Binary Tree. The graph on the left (A) illustrates adding three new nodes to a binary tree, where each new node is connected by a single edge. The table on the right (B) enumerates the connected 3-to-4-node subgraph-to-subgraph transitions (SSTs) caused by adding the new nodes. Columns $u$, $v$, and $w$ depict the number of SSTs in which each corresponding node in (A) is present.}
  \label{fig:bin_tree}
  \vspace{-.1cm}
\end{figure}

When using triadic closure to model graph evolution, the essential question is: ``How many triangles would this new edge close?'' Put in the language of SSTs, we would ask, ``How many wedge-to-triangle transitions would this new edge cause?'' Once we move beyond a single SST such as wedge-to-triangle and into the world of multiple SSTs, we can then observe that any change to a graph corresponds to a collection (i.e. a multiset) of SSTs. Consider the example of a growing binary tree illustrated in Fig.~\ref{fig:bin_tree}. Here the table (B) on the right shows the 3-to-4-node SSTs caused by the addition of a new node and edge. Each addition has its own ``feature vector'' of associated SST counts, {\em i.e.}, the number of SSTs in which each new node and edge is present. In this small example, it is quickly evident that node $v$'s connection, which does not follow the binary-nature of the rest of the graph, has a distinct vector from nodes $u$ and $w$. Considering these collections grants new power for graph modeling, because it grants an important multidimensionality.

The larger the subgraphs one considers for SSTs, the more information one acquires (information-theoretically - meaningful human comprehension may decrease). The largest possible ``subgraph'' to subgraph transition one could consider simply consists of the full old graph and the full new graph.

In the present work we define a model for SSTs on directed and undirected graphs both with and without node and/or edge properties. We provide several analyses to show that SSTs can be used to model graph evolution and static graphs in a variety of contexts simply by fitting linear SVM models to the SST count vectors. Notably, these models are very \textit{interpretable}, yet they perform comparably to state of the art neural network models (with their default hyperparameters) on static and temporal link prediction tasks. We also demonstrate, via a short case study, SSTs intuitively modeling a known graph process.

All our code is available at  \url{https://github.com/SST-Author/Subgraph-Subgraph-Transitions}.

\section{Preliminaries}

Before we formally introduce our SST model we first introduce some preliminary notation. We define a graph in the usual way. A simple directed or undirected graph is defined as $G = (V, E)$, where $V$ is the set of vertices (``nodes'') and $E \subseteq V \times V$ is the set of connections (``edges'').  We use the convention that in undirected graphs $(u, v) = (v, u)$. 

\paragraph{Induced Subgraph}

Given a graph $G = (V, E)$ and a set of nodes $S \subseteq V$, the induced subgraph $G(S)$ is the graph consisting only of the nodes in $S$. Formally, $G(S) = (S, E_{G(S)})$ where $E_{G(S)} = \{(u, v)\ |\ (u, v) \in E \land u, v \in S\}$.

\paragraph{Node and Edge Properties}

A graph's nodes and/or edges may have certain values associated with them. For instance, if an edge indicates a road, it might have a speed limit value. In some cases these properties may be important in how to model the graph. In those cases, we redefine the graph as follows: Let $G = (V, E, p_V, p_E)$ be a graph with two property functions, $p_V$ and $p_E$. Each property function maps a node/edge and a property name to a value. 

\paragraph{Isomorphisms and Automorphism Orbits}\label{sec:iso_and_auto}

Given two graphs $G_1 = (V_1, E_1)$, $G_2 = (V_2, E_2)$ an \textit{isomorphism} is a bijection $f: V_1 \rightarrow V_2$ such that $(u, v) \in E_1 \leftrightarrow (f(u), f(v)) \in E_2$. That is, an isomorphism is a mapping of a graph's nodes to another graph's nodes in a way that lines up the structures exactly. An \textit{automorphism} is simply an isomorphism of a graph with itself.

When using node and edge properties ({\em e.g.} $G_1 = (V_1, E_1, p_{V_1}, p_{E_1})$ and $G_2 = (V_2, E_2, p_{V_2}, p_{E_2})$), an isomorphism must also preserve property values. Formally, $\forall v \in V_1.\ p_{V_1}(v) = p_{V_2}(f(v))$ and $\forall (u, v) \in E_1.\ p_{E_1}((u, v)) = p_{E_2}((f(u), f(v)))$.

The \textit{automorphism orbit} of a node $v \in V$ is the set of nodes to which $v$ is equivalent under automorphism. Formally, $AO(v) = \{u\ |\ u \in V \land \exists \ \text{automorphism}\ f.\ f(v) = u\}$. Edges can have automorphism orbits through a similar definition: $AO((u, v)) = \{(a, b)\ |\ (a, b) \in E \land \exists \ \text{automorphism}\ f.\ f(u) = a \land f(v) = b\}$.

Finding automorphism orbits in a graph is frequently thought of in terms of matching or refining a set of ``colors,'' where nodes in the same orbit are given the same color and the original input graph may arbitrarily require that nodes be put in separate orbits by giving them different colors \citep{luks1982isomorphism, mckay2014practical}. We use this idea in our model to find the automorphism orbits of nodes in SSTs with node and/or edge properties.

\section{Subgraph-to-Subgraph Transitions}\label{sec:sst_def}

A subgraph-to-subgraph transition $T$ is defined to be a pair of ``before'' and ``after'' subgraphs:

$$T = (G_T = (V_T, E_T, {p_V}_T, {p_E}_T), G'_T = (V'_T, E'_T, {p_{V'}}_T, {p_{E'}}_T))$$

Thus, there are many, many possible subgraph-to-subgraph transitions (SSTs). In this work, we limit our analyses to incorporate SSTs meeting certain conditions.


Specifically, we focus on modeling edge additions to a graph. Thus we restrict ourselves to the SSTs that indicate the addition of an edge. Additionally, we require that an SST does not include a change in property values. (The only allowed property changes are for a new edge to receive a property value. All existing nodes' or edges' values remain unchanged in the context of the SST.) This restriction on properties is a way to simplify our model/analyses in this work, but conceptually, the SST generalization allows for changing property values as well. We discuss the ways we do allow changing property values during our graph modeling in Section~\ref{sec:prop_update}. Lastly, we require that the ``after'' subgraph be connected.

Notably, the code we release along with this paper includes the ability to acquire SST information for edge deletions, node additions, and node deletions. However, only edge additions are studied in the present work.

\subsection{Properties in SSTs}\label{sec:sst_properties}

We use SSTs to create interpretable models of graph evolution. To do so, we associate changes to a graph with SSTs. If we allowed numeric property values, the number of distinct SSTs would explode combinatorically, thereby making interpretation more difficult. To address this issue, our modeling algorithms require that each property be treated as one of the following:

\begin{enumerate}
    \item \textbf{Class Trait:} A class trait is a property which has a (ideally small) set of possible class values, where no ordering on the values is required.
    \item \textbf{Rank Trait:} A rank trait is a property which requires that the possible values are totally ordered (e.g. numbers). When labeling an SST with rank traits, our modeling algorithm does not use the raw values of the property but rather for each SST, the nodes (or edges) are sorted and the ranks of the nodes (or edges) are used rather than the raw property values. So for example, instead of listing raw PageRank values of the nodes in an SST ({\em e.g.} PageRanks: $\langle 0.34, 0.12, 0.12, 0.025 \rangle$), the SST would be encoded just with the relative ordering of those values ({\em e.g.} ``PageRank Ordering: $\langle 2, 1, 1, 0 \rangle$'').
\end{enumerate}

\section{SST Graph Model}

Given our formalism, we implement a graph model that encodes and uses SSTs to model graph evolution. This graph model has three distinct modules:

\begin{itemize}[topsep=0pt]
    \item A ``Transition Labeler,'' which takes a before and after subgraph and produces a canonical ({\em i.e.} automorphism-invariant) label for that subgraph-to-subgraph transition.
    \item A ``Transition Counter,'' which takes a change to a graph (an edge addition, edge deletion, node addition, or node deletion) and enumerates all the SSTs induced by that graph change.
    \item Interpretable Static and Temporal Link Predictors which make use of the Transition Counter information.
\end{itemize}


\subsection{The Transition Labeler}

To correctly identify a subgraph-to-subgraph transition, we need to label it in a way that maps all isomorphically equivalent SSTs to the same label ({\em i.e.} a ``canonical label''). To do this, we use an adaption of the Weisfeiler-Lehman isomorphism algorithm \citep{weisfeiler1968reduction}, (a process also known as ``Color Refinement'') which allows node and edge properties to be incorporated as ``colors'' \citep{grohe2017color, berkholz2017tight}. This algorithm provides a canonical node ordering for the vertices involved in the SST. The Weisfeiler-Lehman algorithm is not a \textit{complete} isomorphism algorithm, but it is guaranteed to work on up to 9-node graphs (SSTs) \citep{leman1970automorphisms}.

As discussed earlier, an SST can be thought of as consisting of a ``before'' subgraph and an ``after'' subgraph. At first glance, it may seem that to produce a label for an SST, we must compute distinct canonical labels for the before and after subgraphs and then combine the labels. However, as discussed in Section~\ref{sec:sst_def}, we limit our algorithms to working with four kinds of graph changes: edge addition, edge deletion, node addition, and node deletion, and we do not include property value changes in our SSTs. Thus, each SST we work with can be described as a \textbf{single} subgraph where the added/deleted edge/node is uniquely marked ({\em i.e.} colored); for an example, revisit Figure~\ref{fig:teaser}.

Thus, at a high level, the transition labeler works as follows:
\begin{enumerate}
    \item Receive as input a graph $G = (V, E, p_V, p_E)$, a set of nodes $S \subseteq V$, and a node or edge $x$ to be added to or deleted from the subgraph induced by $S$. In the case of a node addition, the edges by which the new node initially connects to the network must be included in $G$. Similarly, in the case of a node deletion, the edges incident to the deleted node must be included in $G$.
    \item Uniquely label ({\em i.e.} color) $x$.
    \item For any rank traits (see Section~\ref{sec:sst_properties}), temporarily replace the property values with the relative ranking of those values. For example, (``Node Degrees'': $\langle 30, 4, 12, 4 \rangle$) would be replaced with (``Node Degree Ranks'': $\langle 1, 3, 2, 3 \rangle$).
    \item Convert node and edge trait values into node and edge ``colors.'' Each distinct ``color'' corresponds to a unique combination of trait values. This ensures that nodes (or edges) with different property values will be assigned to different automorphism orbits (see Section \ref{sec:iso_and_auto}). 
    \item Given the above coloring, perform canonical color refinement on $G(S)$ to obtain a canonical node ordering $O$ \citep{grohe2017color, berkholz2017tight}.
    \item Use $O$ to serialize $G(S)$, coupled with the information denoting the added/deleted edge or node. This produces a canonical label string $H$.
    \item Hash string $H$ to output a canonical numeric label.
\end{enumerate}

The computational bottlenecks of the algorithm are sorting the values of any included rank traits and running the color refinement algorithm. If $n = |S|$, $m = E_{G(S)}$, $j$ = the number of node traits (properties) and $k$ = the number of edge traits (properties), the ordering of rank traits and other trait processing can be completed in $O(j\,n \log n + k\,m \log m)$. Similarly, using an algorithm developed by Berkholz et. al. which can produce a canonical stable coloring even for edge-colored graphs, the SST labeler can run its ``augmented Weisfeiler-Lehman'' in $O((n + m) \log n)$ time \citep{berkholz2017tight}. Note that our implementation of color refinement is simpler algorithmically but less efficient  than Berkholz et. al.'s ($O(n^2)$), but typically $n$ is small enough that the difference does not matter.

\subsection{The Transition Counter}

To model a graph change ({\em e.g.} an edge addition), we wish to acquire counts of \textit{all} the SSTs of a given size induced by the change. For the kinds of changes we model (edge addition, edge deletion, node addition, node deletion) all the SSTs will involve a few special nodes and their surrounding regions - one special node in the case of a node addition/deletion (the added/deleted node), two in the case of an edge addition/deletion (the edge's endpoints). We first note the one or two nodes involved in all of the SSTs and then employ a technique known as ``Reverse Search'' to enumerate all $k$-node connected subgraphs involving those nodes, where $k$ is the desired SST size \citep{avis1996reverse}. Lastly, for each of these connected subgraphs, we apply our Transition Labeler to obtain a canonical label for the SST.

At present, we are unaware of any techniques to compute the SSTs more efficiently than enumeration. Complex combinatorial tricks allow computing of three, four, and five-node subgraphs in a graph rapidly \citep{jha2015path, pinar2017escape}. At first glance it may seem that a simple solution to avoid enumeration is to efficiently count the subgraphs before and then after the graph change. However, while this would certainly produce useful information, it would not directly produce SSTs, since to know the SST counts one must know \textit{which} subgraphs turned into \textit{which} subgraphs; recall from Figure~\ref{fig:teaser} that one subgraph can often transition into multiple other subgraphs. Additionally, our model requires that SSTs be allowed to have node and edge property values, but the state-of-the-art subgraph counters operate on property-less graphs. Nonetheless we do expect that future researchers will create quick, combinatorial methods for counting SSTs with node and edge properties, and we hope this paper serves as the spark that ignites that project.

As it is, if we hold the SST subgraph size constant at a value $k$, the runtime of our Transition Counter is effectively equivalent to the number of enumerated $k$-node subgraphs around the changes.

\subsubsection{Trait Updaters}\label{sec:prop_update}

Our Transition Labeler forces property values to be the same in both the ``before'' and ``after'' halves of an SST. However our modeling system can still accommodate changes in property values across time. These changes simply are not directly shown in the SSTs. The Transition Counter allows the user to define ``Trait Updaters'' which can update property values before a set of changes is applied, just before a change's collection of SSTs is given labels, just after a change's SSTs are labeled, and after a full set of changes is applied. These options provide great flexibility, which we utilize in our link predictors (Sections \ref{sec:our_static_predictor} and \ref{sec:our_temporal_predictor}).

\subsection{Interpretable Link Predictors}\label{sec:link_pred_overview}

Finally, to demonstrate the power of SSTs, we use them to create interpretable link predictors. 
Link prediction via subgraphs has been discussed by Juszycyszyn et al \citep{juszczyszyn2011link}, Abuoda et al \citep{abuoda2019link}, and Zhang et al \citep{zhang2017weisfeiler}. Likewise, the topic of ``temporal motifs'' distinct from SSTs has been discussed in Liu et al's survey \citep{liu2020temporal}.

Our predictors train for link prediction by collecting vectors of SST counts which correspond to adding actual/positive edges from training samples and vectors of SST counts which correspond to adding randomly sampled non-edges; then we separate the edges' SST vectors from random non-edges' SST vectors with a simple linear SVM. A linear SVM is certainly not the optimal model for prediction accuracy, but even a simple linear SVM with SSTs as its features performs quite well and, importantly, provides a simple way to interpret its predictions: the unit vector which defines the hyperplane separating real edges from non-edges.

Each component of the direction vector corresponds to a distinct SST. The magnitude of the component indicates the relative importance of the SST in distinguishing between actual edges and randomly sampled non-edges. SSTs with positive component values indicate an edge is more likely to be real; SSTs with negative component values indicate an edge is more likely to be a randomly sampled non-edge. We provide examples of interpreting SVM output in the results section.

\subsubsection{Static Link Predictor}\label{sec:our_static_predictor}

A static link predictor is given a single graph and tries to predict which edges may be missing. While SSTs are implicitly designed to model evolving graphs, we can apply them to static graphs relatively easily. To do this, we imagine each edge in the graph as having been ``just added'' by some temporal process. That is, for each edge, we can ask the question, ``What SSTs would be involved if this edge was \textit{not} present and then was added?'' The imagined temporal process we uncover can then predict missing edges in terms of which edges are ``most likely to be added next.''

Our code ``trains'' on every positive edge plus $\alpha$ times as many randomly-sampled non-edges. In our experiments we set $\alpha = 10$.

In a directed graph, SSTs naturally distinguish between the two endpoints of an added edge by the direction of the new edge. While distinguishing between the two vertices being joined is not necessary, it may add useful information. Thus, if the graph is undirected, we create a ``trait updater'' (see Section~\ref{sec:prop_update}) to allow the SSTs to distinguish between the two nodes being connected; whenever the addition of an undirected edge $(u, v)$ is about to have its associated SSTs counted, this trait updater compares the degrees of $u$ and $v$ and then marks them as being of ``equal degree'' or ``higher/lesser'' degree in order to distinguish them.

\subsubsection{Temporal Link Predictor}\label{sec:our_temporal_predictor}

A temporal link predictor operates over series of interactions (edges) with timestamps. In the context of the present work the interactions are allowed to repeat across timestamps. Our temporal link predictor uses a fraction of its training edges as the ``base graph'' and then computes counts for an equal number of true edges and randomly-sampled non-edges.

To make use of the fact that edges have timestamps and may repeat, we create two edge traits and corresponding trait updaters (see Sections \ref{sec:sst_properties} and \ref{sec:prop_update}) to reflect the \textit{recency} and \textit{frequency} of interactions.

The \textit{recency} trait is a ``class trait'' (see Section \ref{sec:sst_properties}) and indicates when an edge most recently occurred. It sorts edges into four categories based on whether the edge:

\begin{enumerate}[topsep=0pt, noitemsep]
    \item has never occurred before (``never'').
    \item last occurred in the previous timestamp (``newest'').
    \item last occurred in the timestamp before the previous (``new'').
    \item last occurred at least three timestamps ago (``old'').
\end{enumerate}

Similarly, the \textit{frequency} trait is also a class trait that sorts edges into four categories based on whether the edge:

\begin{enumerate}[topsep=0pt, noitemsep]
    \item has never occurred before (``0'').
    \item has occurred once before (``1'').
    \item has occurred twice before (``2'').
    \item has occurred three or more times before (``3+'').
\end{enumerate}

These traits for edges allow the SSTs to carry meaning that is simultaneously structural and temporal.

In our temporal link prediction tests (Section \ref{sec:temporal_tests}), the training/validation data is bucketed into nine timestamps. During testing our model uses the first eight timestamps' worth of interactions as the ``base graph'' and then computes the SST vectors for the ninth timestamp. Just using the latest edges for training has a twofold benefit: The edges being trained on and the graph at time of training most closely resemble the edges/graph at test time, and using only the latest edges speeds up training.

\section{Results}

\subsection{Modeling Known Graph Generators}

Before proceeding to show our models operating on real-world graphs, we offer the reader a ``warm-up'' by demonstrating the ability of three-node SSTs to capture the well-known preferential attachment graph generation process first introduced by Barabasi and Albert \citep{barabasi1999emergence}. The preferential attachment model generates a graph by creating a new node and wiring it to $m$ existing nodes, with higher odds of connecting to a node that already has many edges.

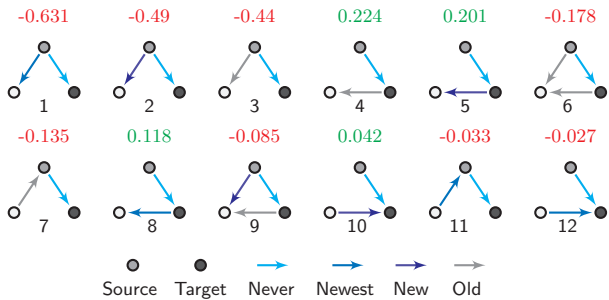
\begin{figure}[tb]
    \centering
    \input{figures/ba_sst}
    \caption{Top 12 3-Node SSTs from a preferential attachment graph process, listed in order of decreasing importance (see Section~\ref{sec:link_pred_overview}). SSTs are combined into a single subgraph where the new edge's source and target nodes are highlighted in gray and dark-gray respectively.
    }
    \label{fig:ba_ssts}
\end{figure}

We generate an example preferential attachment graph with $n=1000$ and $m=2$ and run the temporal link predictor on the temporal sequence of edge additions. As discussed earlier (Section~\ref{sec:link_pred_overview}), the SVM yields weights for each SST, which we use to describe the importance of each SST to the link prediction task. From these SSTs we see that our model captures many key aspects of the preferential attachment process. We illustrate our predictor's top twelve SSTs in Figure~\ref{fig:ba_ssts}, in which the two subgraphs of an SST are combined into a single subgraph where source and target nodes of the new edge are indicated by shaded nodes, and edge colors indicate their recency. A positive weight above the subgraph indicates that the SST is more likely to be associated with a real edge addition than a random edge addition. A negative weight indicates the opposite.

The results indicated in this example are in line with our expectations. The link predictor's top three most important SSTs along with SSTs 6 and 9 all indicate that a node cannot acquire out-edges at distinct timestamps. SSTs 4, 5, and 8 indicate that nodes are pointed-to only after they first point to other nodes, which is a key aspect of the preferential attachment process. Likewise, SST 10 suggests that a node has a higher chance of being pointed to if it was already pointed to recently. Similarly, SSTs 7 and 11 suggest that a node will not begin to point to another node after it has been pointed at. The ordering of SSTs 4, 5, and 8 indicates that newly-formed edges are more likely than randomly-sampled non-edges to point to nodes with older edges; this in turn indicates that nodes with older out-edges have more in-edges. Finally, the relative lack of triangles in the top 12 SSTs suggests that triangles are either rare, uninformative, or both.

These results demonstrate the interpretable power of SSTs to capture a well-known graph generation process which does not follow triadic closure. We now proceed to analyses of real-world graphs, generating both quantitative and interpretable results from the same model.

\subsection{Quantitative Link Prediction Metrics}

Many different metrics are used to quantitatively measure the link prediction performance. Some of the most common are the area under the ROC curve (i.e. ``AUROC'' or just ``AUC''), and Hits at K.

However, Yang et. al. argue that AUC may not be a particularly meaningful metric for link prediction, and Hits at K can provide a very different picture depending on the selected K \citep{yang2015evaluating}. Instead, Yang et. al. demonstrate that area under the precision recall curve (AUPR) may be the best metric both in terms of what it represents and its discriminatory power. Ultimately a model with both high precision and high recall (and thus high AUPR) is of great use, but in an imbalanced setting like link prediction, a model with even a very small false positive \textit{rate} and high true positive rate (and thus a high AUC) can still produce a high \textit{number} of false positives compared to the number of true positives it produces, rendering its link predictions of little use in a real-world setting.

\textit{Unfortunately,} for area under the precision recall curve (AUPR) to be meaningful, negative test cases must not be downsampled \citep{yang2015evaluating}. 
However, having a model score every possible non-existent edge can be quite time-consuming. Thus, rather than reporting link prediction results for a whole graph, Yang et. al. recommend evaluating on the smaller task of link prediction between nodes a max distance of $k$ apart, where $k$ is some small number.

For comparability to other work, we report the AUC. For greater correctness, we report AUPR computed on the limited task of scoring all disconnected pairs of nodes (non-edges) within 3 hops and all connected node pairs (edges) that would be within 3 hops if they were to be disconnected. We call this ``AUPR$_3$'' to differentiate.

\paragraph{Properly Calculating AUPR Curve Areas}

Area under the precision recall curve is often calculated via the trapezoidal rule, which effectively performs a linear interpolation between precision-recall points. This is incorrect, as explained by Davis and Goadrich \citep{davis2006relationship}, who introduce a superior interpolation in their seminal work. This difference becomes particularly relevant when models have large ``gaps'' in their precision recall curves. 

\subsection{Static Link Prediction}

Next, we perform a quantitative and qualitative evaluation on three popular networks, detailed in Table~\ref{tab:static_datasets}, which are frequently used for static link prediction.
Eu-core Emails is a correspondence graph from a European research institute where an edge indicates email(s) sent between two researchers.
Cora ML and Citeseer are paper citation networks where edges indicate citations between papers.

\begin{table}[H]
\centering
\caption{Datasets for link prediction.}
\vspace{-.2cm}
\label{tab:static_datasets}
\small{
\begin{tabular}{@{}c l rrr @{}}
    \hline
    &\multicolumn{1}{c}{\textbf{Dataset}} & \textbf{Node Count} & \multicolumn{1}{c}{\textbf{Edge Count}} & \multicolumn{1}{p{1.5cm}}{\textbf{Temporal Edge Count}} \\ \hline
    \multirow{3}{*}{\rotatebox{90}{\textbf{\tiny{Static}}}} & Eu-core Emails          & 1,005        & 16,706 & --   \\
    & Cora ML                 & 2,708        & 5,278 & --   \\
    & Citeseer                & 3,264        & 45,536 & --   \\
    \addlinespace[0.25em]\hline
    \addlinespace[0.25em]
    \multirow{3}{*}{\rotatebox{90}{\textbf{\tiny{Temporal}}}} & Eu-core Temporal        & 986          & 24,929          & 332,334      \\
    & College Messages         & 1,899         & 20,296          & 59,835       \\
    & Wikipedia               & 100,312      & 746,086         & 1,627,472    \\
    \addlinespace[0.5em]
    \hline
\end{tabular}
}
\end{table}

\begin{table*}[tb]
    \centering
    \caption{Link prediction performance on the static undirected graphs. The best and second-best performing models are  boldfaced and underlined respectively.}
    \input{figures/auc_aupr_table_u}

    \label{tab:static_undir}
\end{table*}

\begin{table*}[tb]
    \centering
    \caption{Link prediction performance on the static directed graphs. The best and second-best performing models are  boldfaced and underlined respectively.}
    \input{figures/auc_aupr_table_d}

    \label{tab:static_dir}
\end{table*}

\begin{table*}[tb]
    \centering
    \caption{Link prediction performance on the temporal directed graphs. The best and second-best performing models are  boldfaced and underlined respectively.}
    \label{tab:temporal_table}
    \input{figures/auc_aupr_table_temporal}
\end{table*}

We use an 85\%/5\%/10\% split of the edges for training, validation, and testing respectively. Because there are no timestamps, the edges are partitioned randomly.


We report results for our models with both 3-node and 4-node SSTs. We compare against 2 baseline models, 4 state-of-the-art graph neural networks (GNNs) for undirected link prediction, and 1 state-of-the-art GNN for directed link prediction. For all the GNNs, we used the default hyperparameters from their source code.

\paragraph{Baseline Models}

We defined two naive baseline methods: random and common neighbor count. The random baseline assigns edge predictions at random. The common neighbors method predicts that the more neighbors two nodes share in common, the more likely those two nodes are to connect \citep{liben2007link}.
Since the common neighbors heuristic does not directly apply to directed graphs, we count each of the four possible directed wedges connecting two nodes for directed graphs, similar to Yin et. al. \citep{yin2019measuring}.

\paragraph{Graph Variational Autoencoders}

Graph Variational Autoencoders (GAEs) \cite{kipf2016variational} have been recently developed to perform deep learning on graphs in support of tasks like link prediction and graph generation.
GAEs are comprised of two parts. First an encoder that embeds a graph into a latent space by applying convolutional layers to an adjacency matrix.
Second, using a simple inner-product decoder, GAEs produce an adjacency matrix of the same dimensions as the original input, which can be used for generating a new graph or for evaluating link prediction on the original graph.

\paragraph{Linear Variational Autoencoders}
In response to the introduction of GAEs, Salha et. al. \cite{salha2020simple} questioned whether convolutional layers are really necessary for performing high-quality node embeddings.
Their proposed Linear Variational Autoencoders (LinearAEs) replace the convolutional layers in GAEs with a simpler one-hop linear model which performs competitively on static link prediction.
The overall behavior is similar to GAEs in that LinearAEs embed a graph's nodes and an inner-product decoder produces a new adjacency matrix for evaluation.

\paragraph{Gravity Graph Variational Autoencoders}
A limitation of both GAEs and LinearAEs lies in their reliance on using inner products of vectors in the latent space for decoding.
This imposes a strong restriction on the decoded adjacency matrices, which must always be symmetric.
To circumvent this limitation, with the goal of performing directed link prediction, Salha et. al. \cite{salha2019gravity} also introduced Gravity-Inspired Graph Variational Autoencoders (GravityAE), capable of generating non-symmetric adjacency matrices using a decoder based on taking sigmoid-activated logarithms of transformed latent vectors.

\subsubsection{Quantitative Results}

The undirected and directed static link prediction results are detailed in Tables \ref{tab:static_undir} and \ref{tab:static_dir} respectively.
The SST-based models are consistently among the top performers.

It is important to note that the GNNs were trained with their default hyperparameters; no hyperparameter optimization was performed. This should make us take the GNNs' lower performance relative to our SST models' with a grain of salt, as our models have the advantage of requiring almost no hyperparameter tuning.

The key takeaway is \textit{not} that our SST models will provide the best link prediction scores. Rather, the takeaway is that they provide \textit{good} quantitative performance, \textit{and thus our models' elegant and interpretable results are valid}.

\subsubsection{Interpretation}\label{sec:cora_ml_intro}

As a case-study in the interpretability of SSTs on real-world graphs, we analyze the four-node SSTs from the Cora ML paper citation graph. Recall that the SVM effectively orders SSTs by how strongly they indicate that an edge is either a genuine edge or a randomly-sampled non-edge (Section~\ref{sec:link_pred_overview}).

\begin{figure}[tb]
    \centering
    \input{figures/cora_only_bidir}
    \caption{Top Cora SSTs with Bidirected Citations -- (See Section~\ref{sec:cora_ml_intro}) -- SSTs 2, 3, and 6 indicate that if articles $A$ and $B$ mutually cite each other, $A$ tends to cite whatever $B$ cites \textit{unless} another article $C$ who bi-cites with $B$ does not. SSTs 4 and 5 indicate that articles are more likely to bi-cite each other if they cite the same articles.}
    \label{fig:cora_sst_bidir}
    \vspace{.2cm}
\end{figure}
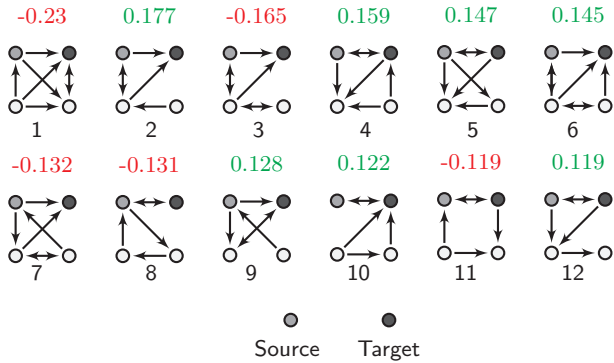

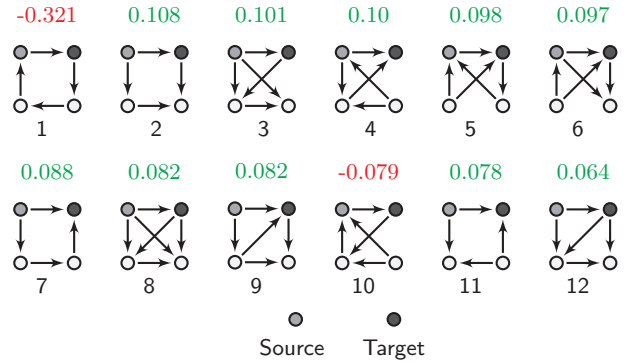
\begin{figure}[tb]
    \centering
    \input{figures/cora_new_sst}
    \caption{Top Cora ML SSTs Without Bidirected Citations -- (See Section~\ref{sec:cora_ml_intro}) -- The top SST indicates that if an edge closes a 4-cycle that is considered a strong indicator that the edge is not genuine. Similarly, SST 10 suggests that a 3-cycle is unlikely, but not as unlikely as a 4-cycle. Other than SSTs 1 and 4, the top SSTs are positive indicators. SSTs 2 - 9, and 11 - 12 all include some kind of ``transitivity'', that nodes which cite (or are cited by) similar articles cite each other.}
    \label{fig:cora_sst_no_bidir}
\end{figure}

We find that the SSTs ranked highest tend to involve bidirected edges (papers that cite each other, perhaps via pre-prints). Sometimes these SSTs are used to predict the presence/non-presence of bidirected edges; sometimes they simply use nearby bidirected edges as indicators of single-direction links. Predicting when a bidirected citation edge forms is a fascinating and difficult task but has limited applicability (Only 2.8\% of connections in the Cora ML graph are bidirected.). Remember that the SVM ranks SSTs by how informative they are \textit{if or when} they occur - \textit{not} by how often they occur. Thus we look at both the SVM's top SSTs \textit{with} bidirected edges and (to get a sense for how the SVM ranks more frequent SSTs) the top SSTs \textit{without} bidirected edges. These are depicted and analyzed in Figures \ref{fig:cora_sst_bidir} and \ref{fig:cora_sst_no_bidir} respectively. The SSTs pick up intuitive aspects of a citation network as well as some intriguing results.

\subsection{Temporal Network Evolution}\label{sec:temporal_tests}

For evaluating networks' behavior over time, we perform future link prediction on three topologically rich, dynamic datasets, summarized in Table~\ref{tab:static_datasets}.
Eu-core Temporal is a time-attributed version of the earlier Eu-core Emails dataset, incorporating timestamps on the  emails.
College Messages is a dynamic social network where edges indicate messages between users at certain times.
Wikipedia is a temporal hyperlink network where the addition of a hyperlink from one page to another is represented by a timestamped edge.



For each network, the edges at time $t$ indicate \textbf{interactions} at time $t$ that can then be repeated at a later time.

Similar to the methodology used by Kasat et al, we bucket the interactions into $\tau$ evenly-sized buckets \citep{kasattemporal}. Since a bucket may cover multiple timestamps, an interaction (edge) may occur multiple times in a single bucket. We squash these multiple occurrences into a single edge and weight the interaction by its number of occurrences. In each bucket an edge's original timestamp is replaced with the index of that bucket. Thus we effectively have a series of $\tau$ graphs, $G_1, ..., G_\tau$. We train on the first $\tau - 1$ and test on $G_\tau$. In our experiments we set $\tau = 10$. Note that neither our model nor the models we compare against make use of the weights; they just use the topology and the timestamps. However, if desired, one could add a ``class trait'' or ``rank trait'' (Section~\ref{sec:sst_properties}) to our temporal link predictor allowing it to make use of these values.

Temporal link predictors are fewer in number than their static counterparts. We compare against one state-of-the-art graph neural network and the baselines from before. As in our static evaluation, we test with both three-node SSTs and four-node SSTs.

\subsubsection{Temporal Graph Neural Networks}

As a state-of-the-art baseline for comparison on the task of temporal link prediction, we rely on the Temporal Graph Networks (TGNs) introduced by Rossi et. al. \citep{rossi2020temporal}.
Their TGN is a graph autoencoder capable of temporally embedding a sequence of events on a graph (e.g., node additions or deletions) using temporal graph attention layers.
A Multi-Layer Perceptron decoder allows the TGN to score candidate edges with probabilities for evaluation of future link prediction.




\subsubsection{Quantitative Results}

Quantitative results are listed in Table~\ref{tab:temporal_table}. Once again our SST-based link predictors are among the top performers. Again, we suggest that these numbers be taken with a grain of salt because we simply used the GNNs' default hyperparameters. Chiefly, our tests demonstrate that our SSTs' elegant and interpretable results are validated by good prediction performance.

Note that we bypassed computing AUPR$_3$ on the Wikipedia graph due to the sheer size of the false test edge set - $O((10^5)^2)$.

\begin{figure}[tb]
    \centering
    \input{figures/wiki_sst}
    \caption{Top 3-Node SSTs for Wikipedia Link Additions -- Main Takeaway: Wedges only close to triangles when the wedge had recent edges ({\em e.g.} Newest) appearing for the first or maybe second time (low frequency, {\em e.g.} `1'), ideally including an edge pointing to the target node of the new edge.}
    \label{fig:wiki_ssts}
\end{figure}
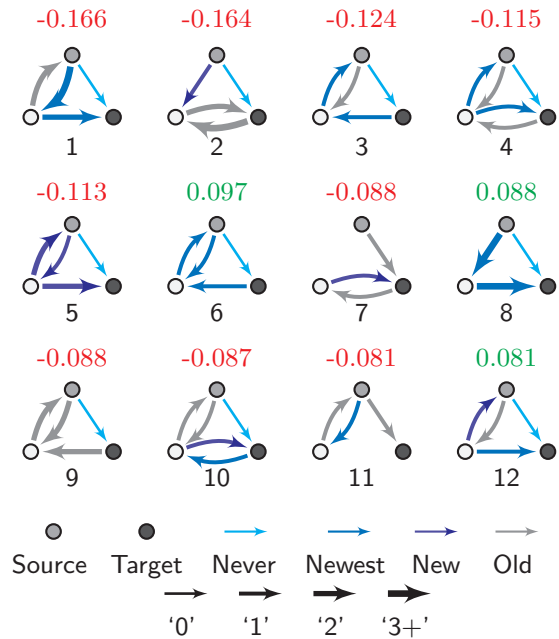

\subsubsection{Interpretable Temporal Results}

To demonstrate the interpretability of SSTs on temporal graphs, we explore the three-node SSTs on the Wikipedia edge additions graph. We find that, unlike the general assumption of triadic closure, according to our model many triangles are considered \textit{unlikely} to close. It is only the triangles where certain connection combinations in the wedge were formed recently (indicated by our recency trait) and \textit{for the first (or maybe second) time} (indicated by our frequency trait) that the wedge is quite likely to close into a triangle. See Figure~\ref{fig:wiki_ssts}. This is evidenced quantitatively by the fact that the three-node SST predictor performed much better than the Common Neighbors.

\section{Conclusion}

We defined an elegant generalization of Triadic Closure, the Subgraph-to-Subgraph Transition (SST). This generalization allowed us to use a simple classifier, the Linear SVM, to create interpretable link prediction models which performed comparatively with state of the art graph neural networks. We expect that the Subgraph-to-Subgraph Transition will become a standard tool in modeling graphs and that future research will produce new and creative ways to use and efficiently count SSTs.

\section*{Acknowledgements}
This research is supported by a grant from the US National Science Foundation (\#1652492). 

\newpage

%% file: figures/fig1.tex
\begin{tikzpicture}[scale=0.825, transform shape]
\begin{scope}[scale=0.75, transform shape, shift={(-1.5,0)}]
	\begin{scope}
		\node [inode] (v1) at (-4.75,4.25) {};
		\node [inode] (v2) at (-4,4.25) {};
		\node [inode] (v3) at (-4,3.5) {};
		\node [inode] (v4) at (-4.75,3.5) {};
		\node [blob] (v15) at (-4.375,3.875) {};
	\end{scope}
	
	\begin{scope}[shift={(2,0)}]
		\node [inode] (v1) at (-4.75,4.25) {};
		\node [inode] (v2) at (-4,4.25) {};
		\node [inode] (v3) at (-4,3.5) {};
		\node [inode] (v4) at (-4.75,3.5) {};
		\node [blob] (v14) at (-4.375,3.875) {};
		
		\draw [iedge] (v4) edge (v1);
	\end{scope}
	
	\begin{scope}[shift={(4,0.75)}]
		\node [inode] (v1) at (-4.75,4.25) {};
		\node [inode] (v2) at (-4,4.25) {};
		\node [inode] (v3) at (-4,3.5) {};
		\node [inode] (v4) at (-4.75,3.5) {};
		\node [blob] (v12) at (-4.375,3.875) {};
		
		\draw [iedge] (v3) edge (v4);
		\draw [iedge] (v4) edge (v1);
	\end{scope}
	
	\begin{scope}[shift={(4,-0.75)}]
		\node [inode] (v1) at (-4.75,4.25) {};
		\node [inode] (v2) at (-4,4.25) {};
		\node [inode] (v3) at (-4,3.5) {};
		\node [inode] (v4) at (-4.75,3.5) {};
		\node [blob] (v13) at (-4.375,3.875) {};
	
		\draw [iedge] (v4) edge (v1);
		\draw [iedge] (v3) edge (v2);
	\end{scope}
	
	\begin{scope}[shift={(6,1.25)}]
		\node [inode] (v1) at (-4.75,4.25) {};
		\node [inode] (v2) at (-4,4.25) {};
		\node [inode] (v3) at (-4,3.5) {};
		\node [inode] (v4) at (-4.75,3.5) {};
		\node [blob] (v9) at (-4.375,3.875) {};
		
		\draw [iedge] (v4) edge (v2);
		\draw [iedge] (v3) edge (v4);
		\draw [iedge] (v4) edge (v1);
	\end{scope}
	
	\begin{scope}[shift={(6,-0.25)}]
		\node [inode] (v1) at (-4.75,4.25) {};
		\node [inode] (v2) at (-4,4.25) {};
		\node [inode] (v3) at (-4,3.5) {};
		\node [inode] (v4) at (-4.75,3.5) {};
		\node [blob] (v10) at (-4.375,3.875) {};
		
		\draw [iedge] (v3) edge (v4);
		\draw [iedge] (v4) edge (v1);
		\draw [iedge] (v1) edge (v3);
	\end{scope}
	
	\begin{scope}[shift={(6,-1.75)}]
		\node [inode] (v1) at (-4.75,4.25) {};
		\node [inode] (v2) at (-4,4.25) {};
		\node [inode] (v3) at (-4,3.5) {};
		\node [inode] (v4) at (-4.75,3.5) {};
		\node [blob] (v11) at (-4.375,3.875) {};
		
		\draw [iedge] (v2) edge (v3);
		\draw [iedge] (v3) edge (v4);
		\draw [iedge] (v1) edge (v4);
	\end{scope}
	
	\begin{scope}[shift={(8,0.75)}]
		\node [inode] (v1) at (-4.75,4.25) {};
		\node [inode] (v2) at (-4,4.25) {};
		\node [inode] (v3) at (-4,3.5) {};
		\node [inode] (v4) at (-4.75,3.5) {};
		\node [blob] (v7) at (-4.375,3.875) {};
		
		\draw [iedge] (v2) edge (v3);
		\draw [iedge] (v3) edge (v4);
		\draw [iedge] (v4) edge (v1);
		\draw [iedge] (v4) edge (v2);
	\end{scope}
	
	\begin{scope}[shift={(8,-0.75)}]
		\node [inode] (v1) at (-4.75,4.25) {};
		\node [inode] (v2) at (-4,4.25) {};
		\node [inode] (v3) at (-4,3.5) {};
		\node [inode] (v4) at (-4.75,3.5) {};
		\node [blob] (v8) at (-4.375,3.875) {};
		
		\draw [iedge] (v1) edge (v2);
		\draw [iedge] (v2) edge (v3);
		\draw [iedge] (v3) edge (v4);
		\draw [iedge] (v4) edge (v1);
	\end{scope}
	
	\begin{scope}[shift={(10,0)}]
		\node [inode] (v1) at (-4.75,4.25) {};
		\node [inode] (v2) at (-4,4.25) {};
		\node [inode] (v3) at (-4,3.5) {};
		\node [inode] (v4) at (-4.75,3.5) {};
		\node [blob] (v5) at (-4.375,3.875) {};
		
		\draw [iedge] (v1) edge (v2);
		\draw [iedge] (v2) edge (v3);
		\draw [iedge] (v3) edge (v4);
		\draw [iedge] (v4) edge (v1);
		\draw [iedge] (v4) edge (v2);
	\end{scope}
	
	\begin{scope}[shift={(12,0)}]
		\node [inode] (v1) at (-4.75,4.25) {};
		\node [inode] (v2) at (-4,4.25) {};
		\node [inode] (v3) at (-4,3.5) {};
		\node [inode] (v4) at (-4.75,3.5) {};
		\node [blob] (v6) at (-4.375,3.875) {};
		
		\draw [iedge] (v1) edge (v2);
		\draw [iedge] (v2) edge (v3);
		\draw [iedge] (v3) edge (v4);
		\draw [iedge] (v4) edge (v1);
		\draw [iedge] (v1) edge (v3);
		\draw [iedge] (v4) edge (v2);
	\end{scope}
	
	\draw [diredge] (v5) edge node [above] {\small\textsf n} (v6);
	\draw [diredge] (v7) edge node [above] {\small\textsf l} (v5);
	\draw [diredge] (v8) edge node [above] {\small\textsf m} (v5);
	\draw [diredge] (v9) edge node [above] {\small\textsf h} (v7);
	\draw [diredge] (v10) edge node [above] {\small\textsf i} (v7);
	\draw [diredge] (v11) edge node [above] {\small\textsf j} (v7);
	\draw [diredge] (v11) edge node [above] {\small\textsf k} (v8);
	\draw [diredge] (v12) edge node [above] {\small\textsf d} (v9);
	\draw [diredge] (v12) edge node [above] {\small\textsf e} (v10);
	\draw [diredge] (v12) edge node [above] {\small\textsf f} (v11);
	\draw [diredge] (v13) edge node [above] {\small\textsf g} (v11);
	\draw [diredge] (v14) edge node [above] {\small\textsf b} (v12);
	\draw [diredge] (v14) edge node [above] {\small\textsf c} (v13);
	\draw [diredge] (v15) edge node [above] {\small\textsf a} (v14);
\end{scope}

\begin{scope}[shift={(6.6515,1)}]
	\begin{scope}[shift={(-5,-5)}]
		\node [inode] (v1) at (-4.75,4.25) {};
		\node [inode] (v2) at (-4,4.25) {};
		\node [inode] (v3) at (-4,3.5) {};
		\node [inode] (v4) at (-4.75,3.5) {};
		
		\draw [dottededge] (v1) edge (v4);
	\end{scope}
	
	\begin{scope}[shift={(-3.5,-5)}]
		\node [inode] (v1) at (-4.75,4.25) {};
		\node [inode] (v2) at (-4,4.25) {};
		\node [inode] (v3) at (-4,3.5) {};
		\node [inode] (v4) at (-4.75,3.5) {};
		
		\draw [iedge] (v4) edge (v1);
		\draw [dottededge] (v4) edge (v3);
	\end{scope}
	
	\begin{scope}[shift={(-2,-5)}]
		\node [inode] (v1) at (-4.75,4.25) {};
		\node [inode] (v2) at (-4,4.25) {};
		\node [inode] (v3) at (-4,3.5) {};
		\node [inode] (v4) at (-4.75,3.5) {};
	
		\draw [iedge] (v4) edge (v1);
		\draw [dottededge] (v3) edge (v2);
	\end{scope}
	
	\begin{scope}[shift={(-0.5,-5)}]
		\node [inode] (v1) at (-4.75,4.25) {};
		\node [inode] (v2) at (-4,4.25) {};
		\node [inode] (v3) at (-4,3.5) {};
		\node [inode] (v4) at (-4.75,3.5) {};
		
		\draw [dottededge] (v4) edge (v2);
		\draw [iedge] (v3) edge (v4);
		\draw [iedge] (v4) edge (v1);
	\end{scope}
	
	\begin{scope}[shift={(1,-5)}]
		\node [inode] (v1) at (-4.75,4.25) {};
		\node [inode] (v2) at (-4,4.25) {};
		\node [inode] (v3) at (-4,3.5) {};
		\node [inode] (v4) at (-4.75,3.5) {};
		
		\draw [dottededge] (v1) edge (v3);
		\draw [iedge] (v3) edge (v4);
		\draw [iedge] (v4) edge (v1);
	\end{scope}
	
	\begin{scope}[shift={(-5,-6.5)}]
		\node [inode] (v1) at (-4.75,4.25) {};
		\node [inode] (v2) at (-4,4.25) {};
		\node [inode] (v3) at (-4,3.5) {};
		\node [inode] (v4) at (-4.75,3.5) {};
		
		\draw [dottededge] (v2) edge (v3);
		\draw [iedge] (v3) edge (v4);
		\draw [iedge] (v4) edge (v1);
	\end{scope}
	
	\begin{scope}[shift={(-3.5,-6.5)}]
		\node [inode] (v1) at (-4.75,4.25) {};
		\node [inode] (v2) at (-4,4.25) {};
		\node [inode] (v3) at (-4,3.5) {};
		\node [inode] (v4) at (-4.75,3.5) {};
		
		\draw [iedge] (v2) edge (v3);
		\draw [dottededge] (v3) edge (v4);
		\draw [iedge] (v4) edge (v1);
	\end{scope}
	
	\begin{scope}[shift={(-2,-6.5)}]
		\node [inode] (v1) at (-4.75,4.25) {};
		\node [inode] (v2) at (-4,4.25) {};
		\node [inode] (v3) at (-4,3.5) {};
		\node [inode] (v4) at (-4.75,3.5) {};
		
		\draw [dottededge] (v2) edge (v3);
		\draw [iedge] (v4) edge (v2);
		\draw [iedge] (v3) edge (v4);
		\draw [iedge] (v4) edge (v1);
	\end{scope}
	
	\begin{scope}[shift={(-0.5,-6.5)}]
		\node [inode] (v1) at (-4.75,4.25) {};
		\node [inode] (v2) at (-4,4.25) {};
		\node [inode] (v3) at (-4,3.5) {};
		\node [inode] (v4) at (-4.75,3.5) {};
		
		\draw [dottededge] (v2) edge (v3);
		\draw [iedge] (v1) edge (v3);
		\draw [iedge] (v3) edge (v4);
		\draw [iedge] (v4) edge (v1);
	\end{scope}
	
	\begin{scope}[shift={(1,-6.5)}]
		\node [inode] (v1) at (-4.75,4.25) {};
		\node [inode] (v2) at (-4,4.25) {};
		\node [inode] (v3) at (-4,3.5) {};
		\node [inode] (v4) at (-4.75,3.5) {};
		
		\draw [dottededge] (v2) edge (v4);
		\draw [iedge] (v2) edge (v3);
		\draw [iedge] (v3) edge (v4);
		\draw [iedge] (v4) edge (v1);
	\end{scope}
	
	\begin{scope}[shift={(-5,-8)}]
		\node [inode] (v1) at (-4.75,4.25) {};
		\node [inode] (v2) at (-4,4.25) {};
		\node [inode] (v3) at (-4,3.5) {};
		\node [inode] (v4) at (-4.75,3.5) {};
		
		\draw [iedge] (v2) edge (v3);
		\draw [dottededge] (v1) edge (v2);
		\draw [iedge] (v4) edge (v1);
		\draw [iedge] (v4) edge (v3);
	\end{scope}
	
	\begin{scope}[shift={(-3.5,-8)}]
		\node [inode] (v1) at (-4.75,4.25) {};
		\node [inode] (v2) at (-4,4.25) {};
		\node [inode] (v3) at (-4,3.5) {};
		\node [inode] (v4) at (-4.75,3.5) {};
		
		\draw [iedge] (v2) edge (v3);
		\draw [dottededge] (v1) edge (v2);
		\draw [iedge] (v4) edge (v1);
		\draw [iedge] (v4) edge (v3);
		\draw [iedge] (v2) edge (v4);
	\end{scope}
	
	\begin{scope}[shift={(-2,-8)}]
		\node [inode] (v1) at (-4.75,4.25) {};
		\node [inode] (v2) at (-4,4.25) {};
		\node [inode] (v3) at (-4,3.5) {};
		\node [inode] (v4) at (-4.75,3.5) {};
		
		\draw [iedge] (v2) edge (v3);
		\draw [iedge] (v1) edge (v2);
		\draw [iedge] (v4) edge (v1);
		\draw [iedge] (v4) edge (v3);
		\draw [dottededge] (v2) edge (v4);
	\end{scope}
	
	\begin{scope}[shift={(-0.5,-8)}]
		\node [inode] (v1) at (-4.75,4.25) {};
		\node [inode] (v2) at (-4,4.25) {};
		\node [inode] (v3) at (-4,3.5) {};
		\node [inode] (v4) at (-4.75,3.5) {};
		
		\draw [iedge] (v1) edge (v2);
		\draw [iedge] (v2) edge (v3);
		\draw [dottededge] (v1) edge (v3);
		\draw [iedge] (v4) edge (v1);
		\draw [iedge] (v4) edge (v3);
		\draw [iedge] (v2) edge (v4);
	\end{scope}
	\node [textnode] at (-9.35,-1.75) {\small\textsf a};
	\node [textnode] at (-7.85,-1.75) {\small\textsf b};
	\node [textnode] at (-6.35,-1.75) {\small\textsf c};
	\node [textnode] at (-4.85,-1.75) {\small\textsf d};
	\node [textnode] at (-3.35,-1.75) {\small\textsf e};
	
	\node [textnode] at (-9.35,-3.25) {\small\textsf f};
	\node [textnode] at (-7.85,-3.25) {\small\textsf g};
	\node [textnode] at (-6.35,-3.25) {\small\textsf h};
	\node [textnode] at (-4.85,-3.25) {\small\textsf i};
	\node [textnode] at (-3.35,-3.25) {\small\textsf j};
	
	\node [textnode] at (-9.35,-4.75) {\small\textsf k};
	\node [textnode] at (-7.85,-4.75) {\small\textsf l};
	\node [textnode] at (-6.35,-4.75) {\small\textsf m};
	\node [textnode] at (-4.85,-4.75) {\small\textsf n};
\end{scope}
\node at (-4.75,4.75) {\large\textsf{\textbf{A}}};
\node at (-5,0.5) {\large\textsf{\textbf{B}}};
\end{tikzpicture}

%% file: figures/fig2.tex
\begin{tikzpicture}[scale=0.875, transform shape]

\begin{scope}[shift={(0.2332,-1.65)}]
	\node [inode] (v1) at (0,0) {};
	\node [inode] (v2) at (-0.5,-0.5) {};
	\node [inode] (v8) at (0.5,-0.5) {};
	\node [inode] (v3) at (-1,-1) {};
	\node [inode, fill=blue!80!cyan] (v5) at (-0.25,-1) {};
	\node [inode] (v9) at (0.25,-1) {};
	\node [inode] (v10) at (0.75,-1) {};
	\node [inode] (v4) at (-1.5,-1.5) {};
	\node [inode] (v6) at (-0.75,-1.5) {};
	\node [inode, fill=blue!80!cyan] (v7) at (-0.5,-2) {};
	\node [inode, fill=red!80!orange] (v13) at (1.25,-1) {};
	\node [inode] (v11) at (0.5,-1.5) {};
	\node [inode] (v12) at (1,-1.5) {};
	
	\draw [diredge] (v1) edge (v2);
	\draw [diredge] (v2) edge (v3);
	\draw [diredge] (v3) edge (v4);
	\draw [diredge, draw=blue!80!cyan] (v2) edge[draw=blue!80!cyan] (v5);
	\draw [diredge] (v3) edge (v6);
	\draw [diredge, draw=blue!80!cyan] (v6) edge[draw=blue!80!cyan] (v7);
	\draw [diredge] (v1) edge (v8);
	\draw [diredge] (v8) edge (v9);
	\draw [diredge] (v8) edge (v10);
	\draw [diredge] (v10) edge (v11);
	\draw [diredge] (v10) edge (v12);
	\draw [diredge, draw=red!80!orange] (v8) edge[draw=red!80!orange] (v13);
	
	\node [textnode] at (-0.4,-1.75) {\small{\textsf{\textcolor{blue!80!cyan}{$w$}}}};
	\node [textnode] at (-0.15,-0.725) {\small{\textsf{\textcolor{blue!80!cyan}{$u$}}}};
	\node [textnode] at (1.25,-0.725) {\small{\textsf{\textcolor{red!80!orange}{$v$}}}};
\end{scope}

\begin{scope}
	\node [inode] (v14) at (3.25,0.75) {};
	\node [inode] (v15) at (3,0.25) {};
	\node [inode] (v16) at (2.75,-0.25) {};
	\draw [diredge] (v14) edge (v15);
	\draw [diredge] (v15) edge (v16);
	\node [inode] (v17) at (4.75,0.75) {};
	\node [inode] (v18) at (4.5,0.25) {};
	\node [inode] (v19) at (4.25,-0.25) {};
	\node [inode] (v20) at (4.75,-0.25) {};
	\draw [diredge] (v17) edge (v18);
	\draw [diredge] (v18) edge (v19);
	\draw [diredge] (v18) edge (v20);
	\node [textnode] at (3.875,0.25) {\large $\Rightarrow$};
\end{scope}

\begin{scope}[shift={(0,-1.5)}]
	\node [inode] (v14) at (3,0.5) {};
	\node [inode] (v15) at (2.75,0) {};
	\node [inode] (v16) at (3.25,0) {};
	\draw [diredge] (v14) edge (v15);
	\draw [diredge] (v14) edge (v16);
	
	\node [textnode] at (3.875,0.25) {\large $\Rightarrow$};
	
	\node [inode] (v17) at (4.75,0.75) {};
	\node [inode] (v18) at (4.5,0.25) {};
	\node [inode] (v19) at (5,0.25) {};
	\node [inode] (v20) at (4.75,-0.25) {};
	\draw [diredge] (v17) edge (v18);
	\draw [diredge] (v17) edge (v19);
	\draw [diredge] (v18) edge (v20);
\end{scope}

\begin{scope}[shift={(0,-3)}]
	\node [inode] (v14) at (3,0.75) {};
	\node [inode] (v15) at (2.75,0.25) {};
	\node [inode] (v16) at (3.25,0.25) {};
	\draw [diredge] (v14) edge (v15);
	\draw [diredge] (v14) edge (v16);
	
	\node [textnode] at (3.875,0.475) {\large $\Rightarrow$};
	
	\node [inode] (v17) at (4.75,0.75) {};
	\node [inode] (v18) at (4.25,0.25) {};
	\node [inode] (v19) at (4.75,0.25) {};
	\node [inode] (v20) at (5.25,0.25) {};
	\draw [diredge] (v17) edge (v18);
	\draw [diredge] (v17) edge (v19);
	\draw [diredge] (v17) edge (v20);
\end{scope}

\begin{scope}[shift={(0,-4)}]
	\node [inode] (v14) at (3.25,0.75) {};
	\node [inode] (v15) at (3,0.25) {};
	\node [inode] (v16) at (2.75,-0.25) {};
	\draw [diredge] (v14) edge (v15);
	\draw [diredge] (v15) edge (v16);
	\node [inode] (v17) at (4.75,0.75) {};
	\node [inode] (v18) at (4.5,0.25) {};
	\node [inode] (v19) at (4.25,-0.25) {};
	\node [inode] (v20) at (5,0.25) {};
	\draw [diredge] (v17) edge (v20);
	\draw [diredge] (v17) edge (v18);
	\draw [diredge] (v18) edge (v19);

	\node [textnode] at (3.875,0.25) {\large $\Rightarrow$};
\end{scope}

\begin{scope}[shift={(0,-5.5)}]
	\node [inode] (v14) at (3.25,0.575) {};
	\node [inode] (v15) at (3,0.075) {};
	\node [inode] (v16) at (3.25,-0.425) {};
	\draw [diredge] (v14) edge (v15);
	\draw [diredge] (v15) edge (v16);
	\node [inode] (v17) at (4.75,0.75) {};
	\node [inode] (v18) at (4.5,0.25) {};
	\node [inode] (v19) at (4.5,-0.75) {};
	\node [inode] (v20) at (4.75,-0.25) {};

	\draw [diredge] (v17) edge (v18);
	\draw [diredge] (v18) edge (v20);
	\draw [diredge] (v20) edge (v19);

	\node [textnode] at (3.875,0.05) {\large $\Rightarrow$};
\end{scope}

\begin{scope}[shift={(0.4836,-0.0556)}]
	\node [textnode] at (7.5,1.5) {\large{\textsf{\textcolor{blue!80!cyan}{$w$}}}};
	\node [textnode] at (6,1.5) {\large{\textsf{\textcolor{blue!80!cyan}{$u$}}}};
	\node [textnode] at (6.75,1.5254) {\large{\textsf{\textcolor{red!80!orange}{$v$}}}};
	
	\node [textnode] at (6,0.25) {\large\textsf 1};
	\node [textnode] at (6.75,0.25) {\large\textsf 2};
	\node [textnode] at (7.5,0.25) {\large\textsf 0};
	
	\node [textnode] at (6,-5.45) {\large\textsf 0};
	\node [textnode] at (6.75,-1.25) {\large\textsf 1};
	\node [textnode] at (7.5,-1.25) {\large\textsf 1};
	
	\node [textnode] at (6,-3.775) {\large\textsf 2};
	\node [textnode] at (6.75,-2.5) {\large\textsf 1};
	\node [textnode] at (7.5,-2.5) {\large\textsf 0};
	
	\node [textnode] at (6,-2.5) {\large\textsf 0};
	\node [textnode] at (6.75,-3.775) {\large\textsf 2};
	\node [textnode] at (7.5,-3.775) {\large\textsf 0};
	
	\node [textnode] at (6,-1.25) {\large\textsf 1};
	\node [textnode] at (6.75,-5.45) {\large\textsf 0};
	\node [textnode] at (7.5,-5.45) {\large\textsf 1};
\end{scope}

\node [textnode] at (-0.75,-1.5) {\large\textsf{\textbf{A}}};
\node [textnode] at (1.5,2) {\large\textsf{\textbf{B}}};

\node [hidden] (v21) at (2.27,1.8332) {};
\node [hidden] (v22) at (8.4,1.8332) {};
\draw [thick]  (v21) edge (v22);  

\node [hidden] (v21) at (2.27,1.166) {};
\node [hidden] (v22) at (8.4,1.166) {};
\draw (v21) edge (v22);  

\node [hidden] (v21) at (2.27,-6.5624) {};
\node [hidden] (v22) at (8.4,-6.5624) {};
\draw [thick]  (v21) edge (v22);  

\node [textnode] at (3.9444,1.4444) {\large\textsf{SST}};
\end{tikzpicture}

%% file: figures/ba_sst.tex
\begin{tikzpicture}[scale=0.8, transform shape]

\begin{scope}[local bounding box=bb]  
	\node [sourceinode] (v1) at (0,0) {};
	\node [inode] (v2) at (-0.5,-0.75) {};
	\node [targetinode] (v3) at (0.5,-0.75) {};
	
	\draw [newestdiredge] (v1) edge (v2);
	\draw [neverdiredge] (v1) edge (v3);
	\node [textnode, text=red] at (0,0.5) {-0.631};
\end{scope}

\begin{scope}[local bounding box=bb, shift={(1.75,0)}]  
	\node [sourceinode] (v1) at (0,0) {};
	\node [inode] (v2) at (-0.5,-0.75) {};
	\node [targetinode] (v3) at (0.5,-0.75) {};
	
	\draw [newdiredge] (v1) edge (v2);
	\draw [neverdiredge] (v1) edge (v3);
	\node [textnode, text=red] at (0,0.5) {-0.49};
\end{scope}

\begin{scope}[local bounding box=bb, shift={(3.5,0)}]  
	\node [sourceinode] (v1) at (0,0) {};
	\node [inode] (v2) at (-0.5,-0.75) {};
	\node [targetinode] (v3) at (0.5,-0.75) {};
	
	\draw [olddiredge] (v1) edge (v2);
	\draw [neverdiredge] (v1) edge (v3);
	\node [textnode, text=red] at (0,0.5) {-0.44};
\end{scope}

\begin{scope}[local bounding box=bb, shift={(5.25,0)}]  
	\node [sourceinode] (v1) at (0,0) {};
	\node [inode] (v2) at (-0.5,-0.75) {};
	\node [targetinode] (v3) at (0.5,-0.75) {};
	
	\draw [neverdiredge] (v1) edge (v3);
	\draw [olddiredge] (v3) edge (v2);
	\node [textnode, text=green] at (0,0.5) {0.224};
\end{scope}

\begin{scope}[local bounding box=bb, shift={(7,0)}]  
	\node [sourceinode] (v1) at (0,0) {};
	\node [inode] (v2) at (-0.5,-0.75) {};
	\node [targetinode] (v3) at (0.5,-0.75) {};
	
	\draw [neverdiredge] (v1) edge (v3);
	\draw [newdiredge] (v3) edge (v2);
	\node [textnode, text=green] at (0,0.5) {0.201};
\end{scope}

\begin{scope}[local bounding box=bb, shift={(8.75,0)}]  
	\node [sourceinode] (v1) at (0,0) {};
	\node [inode] (v2) at (-0.5,-0.75) {};
	\node [targetinode] (v3) at (0.5,-0.75) {};
	
	\draw [olddiredge] (v1) edge (v2);
	\draw [neverdiredge] (v1) edge (v3);
	\draw [olddiredge] (v3) edge (v2);
	\node [textnode, text=red] at (0,0.5) {-0.178};
\end{scope}

\begin{scope}[local bounding box=bb, shift={(0,-2)}]  
	\node [sourceinode] (v1) at (0,0) {};
	\node [inode] (v2) at (-0.5,-0.75) {};
	\node [targetinode] (v3) at (0.5,-0.75) {};
	
	\draw [olddiredge] (v2) edge (v1);
	\draw [neverdiredge] (v1) edge (v3);

	\node [textnode, text=red] at (0,0.5) {-0.135};
\end{scope}

\begin{scope}[local bounding box=bb, shift={(1.75,-2)}]  
	\node [sourceinode] (v1) at (0,0) {};
	\node [inode] (v2) at (-0.5,-0.75) {};
	\node [targetinode] (v3) at (0.5,-0.75) {};
	
	\draw [newestdiredge] (v3) edge (v2);
	\draw [neverdiredge] (v1) edge (v3);

	\node [textnode, text=green] at (0,0.5) {0.118};
\end{scope}

\begin{scope}[local bounding box=bb, shift={(3.5,-2)}]  
	\node [sourceinode] (v1) at (0,0) {};
	\node [inode] (v2) at (-0.5,-0.75) {};
	\node [targetinode] (v3) at (0.5,-0.75) {};
	
	\draw [newdiredge] (v1) edge (v2);
	\draw [neverdiredge] (v1) edge (v3);
	\draw [olddiredge] (v3) edge (v2);
	\node [textnode, text=red] at (0,0.5) {-0.085};
\end{scope}

\begin{scope}[local bounding box=bb, shift={(5.25,-2)}]  
	\node [sourceinode] (v1) at (0,0) {};
	\node [inode] (v2) at (-0.5,-0.75) {};
	\node [targetinode] (v3) at (0.5,-0.75) {};
	
	\draw [newdiredge] (v2) edge (v3);
	\draw [neverdiredge] (v1) edge (v3);

	\node [textnode, text=green] at (0,0.5) {0.042};
\end{scope}

\begin{scope}[local bounding box=bb, shift={(7,-2)}]  
	\node [sourceinode] (v1) at (0,0) {};
	\node [inode] (v2) at (-0.5,-0.75) {};
	\node [targetinode] (v3) at (0.5,-0.75) {};
	
	\draw [neverdiredge] (v1) edge (v3);
	\draw [newestdiredge] (v2) edge (v1);
	\node [textnode, text=red] at (0,0.5) {-0.033};
\end{scope}

\begin{scope}[local bounding box=bb, shift={(8.75,-2)}]  
	\node [sourceinode] (v1) at (0,0) {};
	\node [inode] (v2) at (-0.5,-0.75) {};
	\node [targetinode] (v3) at (0.5,-0.75) {};
	
	\draw [neverdiredge] (v1) edge (v3);
	\draw [newestdiredge] (v2) edge (v3);
	\node [textnode, text=red] at (0,0.5) {-0.027};
\end{scope}

\begin{scope}[shift={(0.7,2.125)}]  
	\node [sourceinode] at (0.775,-5.75) {};
	\node [textnode] at (0.725,-6.175) {\textsf{Source}};
	
	\node [targetinode] at (1.9,-5.75) {};
	\node [textnode] at (1.9,-6.175) {\textsf{Target}};
	
	\node [hidden] (v4) at (2.725,-5.75) {};
	\node [hidden] (v5) at (3.475,-5.75) {};
	\draw [neverdiredge] (v4) edge (v5);
	\node [textnode] at (3.075,-6.175) {\textsf{Never}};
	
	\node [hidden] (v6) at (3.975,-5.75) {};
	\node [hidden] (v7) at (4.725,-5.75) {};
	\draw [newestdiredge] (v6) edge (v7);
	\node [textnode] at (4.3,-6.175) {\textsf{Newest}};
	
	\node [hidden] (v8) at (5.025,-5.75) {};
	\node [hidden] (v9) at (5.775,-5.75) {};
	\draw [newdiredge] (v8) edge (v9);
	\node [textnode] at (5.4,-6.175) {\textsf{New}};
	
	\node [hidden] (v10) at (6,-5.75) {};
	\node [hidden] (v11) at (6.75,-5.75) {};
	\node [textnode] at (6.325,-6.175) {\textsf{Old}};
	\draw [olddiredge] (v10) edge (v11);
\end{scope}

\node [textnode] at (0,-1) {\textsf{1}};
\node [textnode] at (1.75,-1) {\textsf{2}};
\node [textnode] at (3.5,-1) {\textsf{3}};
\node [textnode] at (5.25,-1) {\textsf{4}};
\node [textnode] at (7,-1) {\textsf{5}};
\node [textnode] at (8.7,-1) {\textsf{6}};
\node [textnode] at (0,-3) {\textsf{7}};
\node [textnode] at (1.8,-3) {\textsf{8}};
\node [textnode] at (3.5,-3) {\textsf{9}};
\node [textnode] at (5.2,-3) {\textsf{10}};
\node [textnode] at (6.9,-3) {\textsf{11}};
\node [textnode] at (8.7,-3) {\textsf{12}};
\end{tikzpicture}

%% file: figures/auc_aupr_table_u.tex
\begin{tabular}{lrrcrrcrr}
\hline
    \multirow{2}{*}{\textbf{Model}} & \multicolumn{2}{c}{\textbf{CiteSeer}} &\phantom{}&          \multicolumn{2}{c}{\textbf{Cora ML}} &   \phantom{}&      \multicolumn{2}{c}{\textbf{Eu-core Emails}} \\
    \cmidrule{2-3} \cmidrule{5-6} \cmidrule{8-9}
    
    & AUC   & AUPR$_3$  && AUC   & AUPR$_3$  && AUC   & AUPR$_3$   \\
\hline
CommonNeighbors & 0.669$\pm$0.008 & 0.017$\pm$0.003 &  & 0.716$\pm$0.014 & \underline{0.021}$\pm$0.003 &  & \underline{0.939}$\pm$0.004 & \underline{0.120}$\pm$0.008 \\
GCNAE & 0.784$\pm$0.019 & 0.017$\pm$0.003 &  & 0.847$\pm$0.013 & 0.018$\pm$0.003 &  & 0.912$\pm$0.006 & 0.098$\pm$0.009 \\
GCNVAE & \underline{0.788}$\pm$0.015 & 0.016$\pm$0.002 &  & 0.846$\pm$0.011 & 0.017$\pm$0.003 &  & 0.904$\pm$0.008 & 0.090$\pm$0.013 \\
LinearAE & 0.775$\pm$0.014 & \underline{0.019}$\pm$0.003 &  & 0.829$\pm$0.014 & \underline{0.021}$\pm$0.003 &  & 0.923$\pm$0.005 & \underline{0.120}$\pm$0.005 \\
LinearVAE & 0.786$\pm$0.016 & 0.016$\pm$0.003 &  & \underline{0.848}$\pm$0.017 & 0.019$\pm$0.003 &  & 0.912$\pm$0.005 & 0.106$\pm$0.006 \\
Random & 0.491$\pm$0.019 & 0.004$\pm$0.000 &  & 0.496$\pm$0.018 & 0.002$\pm$0.000 &  & 0.500$\pm$0.012 & 0.004$\pm$0.000 \\
SST-SVM-4 & \textbf{0.865}$\pm$0.017 & \textbf{0.020}$\pm$0.003 &  & \textbf{0.879}$\pm$0.016 & 0.019$\pm$0.004 &  & 0.807$\pm$0.120 & 0.096$\pm$0.033 \\
SST-SVM-3 & 0.754$\pm$0.016 & 0.019$\pm$0.002 &  & 0.823$\pm$0.011 & \textbf{0.024}$\pm$0.003 &  & \textbf{0.943}$\pm$0.004 & \textbf{0.126}$\pm$0.008 \\ \hline

\end{tabular}

%% file: figures/auc_aupr_table_d.tex
\small{
\begin{tabular}{@{}l rr c rr c rr@{}}
\hline
    \multirow{2}{*}{\textbf{Model}} & \multicolumn{2}{c}{\textbf{CiteSeer (D)}} &\phantom{}&          \multicolumn{2}{c}{\textbf{Cora ML (D)}} &   \phantom{}&      \multicolumn{2}{c}{\textbf{Eu-core Emails (D)}} \\
    \cmidrule{2-3} \cmidrule{5-6} \cmidrule{8-9}
    
    & AUC   & AUPR$_3$ && AUC   & AUPR$_3$ && AUC   & AUPR$_3$\\
\hline
CommonNeighbors & 0.669$\pm$0.006 & 0.007$\pm$0.001 &  & 0.721$\pm$0.007 & 0.012$\pm$0.002 &  & \underline{0.947}$\pm$0.002 & 0.103$\pm$0.004 \\
GravityGCNAE & 0.500$\pm$0.013 & 0.002$\pm$0.000 &  & 0.506$\pm$0.015 & 0.001$\pm$0.000 &  & 0.657$\pm$0.018 & 0.004$\pm$0.000 \\
GravityGCNVAE & 0.516$\pm$0.008 & 0.002$\pm$0.000 &  & 0.512$\pm$0.011 & 0.001$\pm$0.000 &  & 0.826$\pm$0.004 & 0.008$\pm$0.000 \\
Random & 0.499$\pm$0.012 & 0.002$\pm$0.000 &  & 0.502$\pm$0.006 & 0.001$\pm$0.000 &  & 0.501$\pm$0.008 & 0.003$\pm$0.000 \\
SST-SVM-4 & \textbf{0.843}$\pm$0.01 & \textbf{0.014}$\pm$0.003 &  & \underline{0.877}$\pm$0.011 & \underline{0.011}$\pm$0.002 &  & 0.886$\pm$0.039 & \underline{0.137}$\pm$0.002 \\
SST-SVM-3 & \underline{0.766}$\pm$0.01 & \underline{0.013}$\pm$0.002 &  & \textbf{0.891}$\pm$0.008 & \textbf{0.018}$\pm$0.002 &  & \textbf{0.970}$\pm$0.001 & \textbf{0.176}$\pm$0.006 \\
\hline
\end{tabular}
}

%% file: figures/auc_aupr_table_temporal.tex
\small{
\begin{tabular}{@{}lrr c rr c rr@{}}
\hline
   \multirow{2}{*}{\textbf{Model}} & \multicolumn{2}{c}{\textbf{College Messages}} & \phantom{} & \multicolumn{2}{c}{\textbf{Eu-core Temporal}} & \phantom{} & \multicolumn{2}{c}{\textbf{Wikipedia}}\\
   \cmidrule{2-3} \cmidrule{5-6} \cmidrule{8-9}
   & AUC   & AUPR$_3$ && AUC  & AUPR$_3$ && AUC & AUPR$_3$ \\
\hline
CommonNeighbors & 0.594$\pm$0.003 & 0.002$\pm$0.000 &  & \textbf{0.938}$\pm$0.001 & \underline{0.201}$\pm$0.000 &  & \underline{0.692}$\pm$0.000 & --- \\
Random & 0.499$\pm$0.007 & 0.001$\pm$0.000 &  & 0.498$\pm$0.004 & 0.008$\pm$0.000 &  & 0.500$\pm$0.001 & --- \\
TGN & \underline{0.749}$\pm$0.000 & \textbf{0.017}$\pm$0.007 &  & 0.762$\pm$0.014 & 0.005$\pm$0.000 &  & --- & --- \\
SST-SVM-4 & 0.669$\pm$0.025 & 0.002$\pm$0.000 &  & 0.89$\pm$0.004 & 0.069$\pm$0.019 &  & --- & --- \\
SST-SVM-3 & \textbf{0.803}$\pm$0.010 & \underline{0.008}$\pm$0.002 &  & \underline{0.933}$\pm$0.003 & \textbf{0.253}$\pm$0.020 &  & \textbf{0.867}$\pm$0.001 & --- \\ \hline

\end{tabular}
}

%% file: figures/cora_only_bidir.tex
\begin{tikzpicture}[scale=0.95, transform shape]

\begin{scope}[local bounding box=bb]  
	\node [sourceinode] (v1) at (-0.5,0) {};
	\node [targetinode] (v2) at (0.25,0) {};
	\node [inode] (v3) at (0.25,-0.75) {};
	\node [inode] (v4) at (-0.5,-0.75) {};
	
	\draw [diredge] (v1) edge (v2);
	\draw [diredge] (v1) edge (v3);
	\draw [diredge, <->] (v2) edge (v3);
	\draw [diredge] (v4) edge (v1);
	\draw [diredge] (v4) edge (v2);
	\draw [diredge] (v4) edge (v3);
	\node [textnode, text=red] at (-0.125,0.5) {-0.23};
\end{scope}

\begin{scope}[local bounding box=bb, shift={(1.5,0)}]  
	\node [sourceinode] (v1) at (-0.5,0) {};
	\node [targetinode] (v2) at (0.25,0) {};
	\node [inode] (v3) at (0.25,-0.75) {};
	\node [inode] (v4) at (-0.5,-0.75) {};
	
	\draw [diredge] (v1) edge (v2);
	\draw [diredge, <->] (v1) edge (v4);
	\draw [diredge] (v3) edge (v4);
	\draw [diredge] (v4) edge (v2);
	\node [textnode, text=green] at (-0.125,0.5) {0.177};
\end{scope}

\begin{scope}[local bounding box=bb, shift={(3,0)}]  
	\node [sourceinode] (v1) at (-0.5,0) {};
	\node [targetinode] (v2) at (0.25,0) {};
	\node [inode] (v3) at (0.25,-0.75) {};
	\node [inode] (v4) at (-0.5,-0.75) {};
	
	\draw [diredge] (v1) edge (v2);
	\draw [diredge, <->] (v1) edge (v4);
	\draw [diredge, <->] (v3) edge (v4);
	\draw [diredge] (v4) edge (v2);
	\node [textnode, text=red] at (-0.125,0.5) {-0.165};
\end{scope}

\begin{scope}[local bounding box=bb, shift={(4.5,0)}]  
	\node [sourceinode] (v1) at (-0.5,0) {};
	\node [targetinode] (v2) at (0.25,0) {};
	\node [inode] (v3) at (0.25,-0.75) {};
	\node [inode] (v4) at (-0.5,-0.75) {};
	
	\draw [diredge, <->] (v1) edge (v2);
	\draw [diredge] (v1) edge (v4);
	\draw [diredge] (v2) edge (v4);
	\draw [diredge] (v3) edge (v2);
	\draw [diredge] (v3) edge (v4);
	\node [textnode, text=green] at (-0.125,0.5) {0.159};
\end{scope}


\begin{scope}[local bounding box=bb, shift={(6,0.01)}]  
	\node [sourceinode] (v1) at (-0.5,0) {};
	\node [targetinode] (v2) at (0.25,0) {};
	\node [inode] (v3) at (0.25,-0.75) {};
	\node [inode] (v4) at (-0.5,-0.75) {};
	
	\draw [diredge, <->] (v1) edge (v2);
	\draw [diredge] (v1) edge (v3);
	\draw [diredge] (v1) edge (v4);
	\draw [diredge] (v2) edge (v4);
	\draw [diredge] (v3) edge (v4);
	\node [textnode, text=green] at (-0.125,0.5) {0.147};
\end{scope}

\begin{scope}[local bounding box=bb, shift={(7.5,0.01)}]  
	\node [sourceinode] (v1) at (-0.5,0) {};
	\node [targetinode] (v2) at (0.25,0) {};
	\node [inode] (v3) at (0.25,-0.75) {};
	\node [inode] (v4) at (-0.5,-0.75) {};
	
	\draw [diredge] (v1) edge (v2);
	\draw [diredge, <->] (v1) edge (v4);
	\draw [diredge] (v3) edge (v2);
	\draw [diredge, <->] (v3) edge (v4);
	\draw [diredge] (v4) edge (v1);
	\draw [diredge] (v4) edge (v2);
	\node [textnode, text=green] at (-0.125,0.5) {0.145};
\end{scope}

\begin{scope}[local bounding box=bb, shift={(0,-2.09)}]  
	\node [sourceinode] (v1) at (-0.5,0) {};
	\node [targetinode] (v2) at (0.25,0) {};
	\node [inode] (v3) at (0.25,-0.75) {};
	\node [inode] (v4) at (-0.5,-0.75) {};
	
	\draw [diredge] (v1) edge (v2);
	\draw [diredge] (v1) edge (v4);
	\draw [diredge] (v3) edge (v1);
	\draw [diredge, <->] (v3) edge (v4);
	\draw [diredge] (v4) edge (v2);
	
	\node [textnode, text=red] at (-0.125,0.5) {-0.132};
\end{scope}

\begin{scope}[local bounding box=bb, shift={(1.5,-2.09)}]  
	\node [sourceinode] (v1) at (-0.5,0) {};
	\node [targetinode] (v2) at (0.25,0) {};
	\node [inode] (v3) at (0.25,-0.75) {};
	\node [inode] (v4) at (-0.5,-0.75) {};
	
	\draw [diredge, <->] (v1) edge (v2);
	\draw [diredge] (v1) edge (v3);
	\draw [diredge] (v3) edge (v4);
	\draw [diredge] (v4) edge (v1);
	
	\node [textnode, text=red] at (-0.125,0.5) {-0.131};
\end{scope}


\begin{scope}[local bounding box=bb, shift={(3,-2.08)}]  
	\node [sourceinode] (v1) at (-0.5,0) {};
	\node [targetinode] (v2) at (0.25,0) {};
	\node [inode] (v3) at (0.25,-0.75) {};
	\node [inode] (v4) at (-0.5,-0.75) {};
	
	\draw [diredge] (v1) edge (v2);
	\draw [diredge] (v1) edge (v4);
	\draw [diredge, <->] (v2) edge (v4);
	\draw [diredge] (v3) edge (v1);
	\node [textnode, text=green] at (-0.125,0.5) {0.128};
\end{scope}

\begin{scope}[local bounding box=bb, shift={(4.5,-2.07)}]  
	\node [sourceinode] (v1) at (-0.5,0) {};
	\node [targetinode] (v2) at (0.25,0) {};
	\node [inode] (v3) at (0.25,-0.75) {};
	\node [inode] (v4) at (-0.5,-0.75) {};
	
	\draw [diredge, <->] (v1) edge (v2);
	\draw [diredge] (v3) edge (v2);
	\draw [diredge] (v4) edge (v2);
	\draw [diredge] (v4) edge (v3);
	
	\node [textnode, text=green] at (-0.125,0.5) {0.122};
\end{scope}

\begin{scope}[local bounding box=bb, shift={(6,-2.05)}]  
	\node [sourceinode] (v1) at (-0.5,0) {};
	\node [targetinode] (v2) at (0.25,0) {};
	\node [inode] (v3) at (0.25,-0.75) {};
	\node [inode] (v4) at (-0.5,-0.75) {};
	
	\draw [diredge, <->] (v1) edge (v2);
	\draw [diredge] (v2) edge (v3);
	\draw [diredge] (v4) edge (v1);
	\draw [diredge] (v4) edge (v3);

	\node [textnode, text=red] at (-0.125,0.5) {-0.119};
\end{scope}

\begin{scope}[local bounding box=bb, shift={(7.5,-2.05)}]  
	\node [sourceinode] (v1) at (-0.5,0) {};
	\node [targetinode] (v2) at (0.25,0) {};
	\node [inode] (v3) at (0.25,-0.75) {};
	\node [inode] (v4) at (-0.5,-0.75) {};
	
	\draw [diredge, <->] (v1) edge (v2);
	\draw [diredge] (v1) edge (v4);
	\draw [diredge] (v2) edge (v4);
	\draw [diredge] (v4) edge (v3);
	\node [textnode, text=green] at (-0.125,0.5) {0.119};
\end{scope}

\begin{scope}[shift={(2.825,2.01)}]  
	\node [sourceinode] at (0.525,-5.75) {};
	\node [textnode] at (0.475,-6.175) {\textsf{Source}};
	
	\node [targetinode] at (1.9,-5.75) {};
	\node [textnode] at (1.9,-6.175) {\textsf{Target}};
\end{scope}

\node [textnode] at (-0.2,-1.1) {\textsf{1}};
\node [textnode] at (1.4,-1.1) {\textsf{2}};
\node [textnode] at (2.9,-1.1) {\textsf{3}};
\node [textnode] at (4.4,-1.1) {\textsf{4}};
\node [textnode] at (5.9,-1.1) {\textsf{5}};
\node [textnode] at (7.3,-1.1) {\textsf{6}};
\node [textnode] at (-0.2,-3.1) {\textsf{7}};
\node [textnode] at (1.4,-3.1) {\textsf{8}};
\node [textnode] at (2.8,-3.1) {\textsf{9}};
\node [textnode] at (4.3,-3.1) {\textsf{10}};
\node [textnode] at (5.8,-3.1) {\textsf{11}};
\node [textnode] at (7.3,-3.1) {\textsf{12}};
\end{tikzpicture}

%% file: figures/cora_new_sst.tex
\begin{tikzpicture}[scale=0.95, transform shape]

\begin{scope}[local bounding box=bb]  
	\node [sourceinode] (v1) at (-0.5,0) {};
	\node [targetinode] (v2) at (0.25,0) {};
	\node [inode] (v3) at (0.25,-0.75) {};
	\node [inode] (v4) at (-0.5,-0.75) {};
	
	\draw [diredge] (v1) edge (v2);
	\draw [diredge] (v2) edge (v3);
	\draw [diredge] (v3) edge (v4);
	\draw [diredge] (v4) edge (v1);
	
	\node [textnode, text=red] at (-0.125,0.5) {-0.321};
\end{scope}

\begin{scope}[local bounding box=bb, shift={(1.5,0)}]  
	\node [sourceinode] (v1) at (-0.5,0) {};
	\node [targetinode] (v2) at (0.25,0) {};
	\node [inode] (v3) at (0.25,-0.75) {};
	\node [inode] (v4) at (-0.5,-0.75) {};
	
	\draw [diredge] (v1) edge (v2);
	\draw [diredge] (v1) edge (v4);
	\draw [diredge] (v2) edge (v3);
	\draw [diredge] (v4) edge (v3);
	\node [textnode, text=green] at (-0.125,0.5) {0.108};
\end{scope}

\begin{scope}[local bounding box=bb, shift={(3,0)}]  
	\node [sourceinode] (v1) at (-0.5,0) {};
	\node [targetinode] (v2) at (0.25,0) {};
	\node [inode] (v3) at (0.25,-0.75) {};
	\node [inode] (v4) at (-0.5,-0.75) {};
	
	\draw [diredge] (v1) edge (v2);
	\draw [diredge] (v1) edge (v3);
	\draw [diredge] (v1) edge (v4);
	\draw [diredge] (v2) edge (v4);
	\draw [diredge] (v4) edge (v3);
	\node [textnode, text=green] at (-0.125,0.5) {0.101};
\end{scope}

\begin{scope}[local bounding box=bb, shift={(4.5,0)}]  
	\node [sourceinode] (v1) at (-0.5,0) {};
	\node [targetinode] (v2) at (0.25,0) {};
	\node [inode] (v3) at (0.25,-0.75) {};
	\node [inode] (v4) at (-0.5,-0.75) {};
	
	\draw [diredge] (v1) edge (v2);
	\draw [diredge] (v1) edge (v4);
	\draw [diredge] (v3) edge (v1);
	\draw [diredge] (v3) edge (v4);
	\draw [diredge] (v4) edge (v2);
	\node [textnode, text=green] at (-0.125,0.5) {0.10};
\end{scope}


\begin{scope}[local bounding box=bb, shift={(6,0)}]  
	\node [sourceinode] (v1) at (-0.5,0) {};
	\node [targetinode] (v2) at (0.25,0) {};
	\node [inode] (v3) at (0.25,-0.75) {};
	\node [inode] (v4) at (-0.5,-0.75) {};
	
	\draw [diredge] (v1) edge (v2);
	\draw [diredge] (v2) edge (v3);
	\draw [diredge] (v3) edge (v1);
	\draw [diredge] (v4) edge (v1);
	\draw [diredge] (v4) edge (v2);
	\node [textnode, text=green] at (-0.125,0.5) {0.098};
\end{scope}

\begin{scope}[local bounding box=bb, shift={(7.5,0)}]  
	\node [sourceinode] (v1) at (-0.5,0) {};
	\node [targetinode] (v2) at (0.25,0) {};
	\node [inode] (v3) at (0.25,-0.75) {};
	\node [inode] (v4) at (-0.5,-0.75) {};
	
	\draw [diredge] (v1) edge (v2);
	\draw [diredge] (v1) edge (v3);
	\draw [diredge] (v2) edge (v3);
	\draw [diredge] (v4) edge (v1);
	\draw [diredge] (v4) edge (v2);
	\node [textnode, text=green] at (-0.125,0.5) {0.097};
\end{scope}

\begin{scope}[local bounding box=bb, shift={(0,-2.2)}]  
	\node [sourceinode] (v1) at (-0.5,0) {};
	\node [targetinode] (v2) at (0.25,0) {};
	\node [inode] (v3) at (0.25,-0.75) {};
	\node [inode] (v4) at (-0.5,-0.75) {};
	
	\draw [diredge] (v1) edge (v2);
	\draw [diredge] (v1) edge (v4);
	\draw [diredge] (v3) edge (v2);
	\draw [diredge] (v4) edge (v3);
	\node [textnode, text=green] at (-0.125,0.5) {0.088};
\end{scope}

\begin{scope}[local bounding box=bb, shift={(1.5,-2.2)}]  
	\node [sourceinode] (v1) at (-0.5,0) {};
	\node [targetinode] (v2) at (0.25,0) {};
	\node [inode] (v3) at (0.25,-0.75) {};
	\node [inode] (v4) at (-0.5,-0.75) {};
	
	\draw [diredge] (v1) edge (v2);
	\draw [diredge] (v1) edge (v3);
	\draw [diredge] (v1) edge (v4);
	\draw [diredge] (v2) edge (v3);
	\draw [diredge] (v2) edge (v4);
	\draw [diredge] (v4) edge (v3);
	\node [textnode, text=green] at (-0.125,0.5) {0.082};
\end{scope}


\begin{scope}[local bounding box=bb, shift={(3,-2.19)}]  
	\node [sourceinode] (v1) at (-0.5,0) {};
	\node [targetinode] (v2) at (0.25,0) {};
	\node [inode] (v3) at (0.25,-0.75) {};
	\node [inode] (v4) at (-0.5,-0.75) {};
	
	\draw [diredge] (v1) edge (v2);
	\draw [diredge] (v1) edge (v4);
	\draw [diredge] (v2) edge (v3);
	\draw [diredge] (v4) edge (v2);
	\draw [diredge] (v4) edge (v3);
	\node [textnode, text=green] at (-0.125,0.5) {0.082};
\end{scope}

\begin{scope}[local bounding box=bb, shift={(4.5,-2.2)}]  
	\node [sourceinode] (v1) at (-0.5,0) {};
	\node [targetinode] (v2) at (0.25,0) {};
	\node [inode] (v3) at (0.25,-0.75) {};
	\node [inode] (v4) at (-0.5,-0.75) {};
	
	\draw [diredge] (v1) edge (v2);
	\draw [diredge] (v2) edge (v4);
	\draw [diredge] (v3) edge (v1);
	\draw [diredge] (v3) edge (v4);
	\draw [diredge] (v4) edge (v1);
	\node [textnode, text=red] at (-0.125,0.5) {-0.079};
\end{scope}

\begin{scope}[local bounding box=bb, shift={(6,-2.2)}]  
	\node [sourceinode] (v1) at (-0.5,0) {};
	\node [targetinode] (v2) at (0.25,0) {};
	\node [inode] (v3) at (0.25,-0.75) {};
	\node [inode] (v4) at (-0.5,-0.75) {};

	\draw [diredge] (v1) edge (v2);
	\draw [diredge] (v1) edge (v4);
	\draw [diredge] (v3) edge (v2);
	\draw [diredge] (v3) edge (v4);
	\node [textnode, text=green] at (-0.125,0.5) {0.078};
\end{scope}

\begin{scope}[local bounding box=bb, shift={(7.5,-2.2)}]  
	\node [sourceinode] (v1) at (-0.5,0) {};
	\node [targetinode] (v2) at (0.25,0) {};
	\node [inode] (v3) at (0.25,-0.75) {};
	\node [inode] (v4) at (-0.5,-0.75) {};
	
	\draw [diredge] (v1) edge (v2);
	\draw [diredge] (v1) edge (v4);
	\draw [diredge] (v2) edge (v3);
	\draw [diredge] (v2) edge (v4);
	\draw [diredge] (v4) edge (v3);
	\node [textnode, text=green] at (-0.125,0.5) {0.064};
\end{scope}

\begin{scope}[shift={(2.825,2.01)}]  
	\node [sourceinode] at (0.525,-5.75) {};
	\node [textnode] at (0.475,-6.175) {\textsf{Source}};
	
	\node [targetinode] at (1.9,-5.75) {};
	\node [textnode] at (1.9,-6.175) {\textsf{Target}};
\end{scope}

\node [textnode] at (-0.2,-1.1) {\textsf{1}};
\node [textnode] at (1.4,-1.1) {\textsf{2}};
\node [textnode] at (2.9,-1.1) {\textsf{3}};
\node [textnode] at (4.4,-1.1) {\textsf{4}};
\node [textnode] at (5.8,-1.1) {\textsf{5}};
\node [textnode] at (7.3,-1.1) {\textsf{6}};
\node [textnode] at (-0.2,-3.3) {\textsf{7}};
\node [textnode] at (1.3,-3.3) {\textsf{8}};
\node [textnode] at (2.8,-3.3) {\textsf{9}};
\node [textnode] at (4.3,-3.3) {\textsf{10}};
\node [textnode] at (5.8,-3.3) {\textsf{11}};
\node [textnode] at (7.3,-3.3) {\textsf{12}};
\end{tikzpicture}

%% file: figures/wiki_sst.tex
\begin{tikzpicture}[scale=1.1, transform shape]

\begin{scope}[local bounding box=bb, shift={(0,0.1365)}]  
	\node [sourceinode] (v1) at (0,0) {};
	\node [targetinode] (v2) at (0.5,-0.75) {};
	\node [inode] (v3) at (-0.5,-0.75) {};
	
	\draw [neverdiredge, t0] (v1) edge (v2);
	\draw [newestdiredge, t3] (v1) edge[bend left=25] (v3);
	\draw [olddiredge, t2] (v3) edge[bend left=25] (v1);
	\draw [newestdiredge, t2] (v3) edge (v2);
	
	\node [textnode, text=red] at (0,0.409) {-0.166};
\end{scope}

\begin{scope}[local bounding box=bb, shift={(1.75,0.1365)}]  
	\node [sourceinode] (v1) at (0,0) {};
	\node [inode] (v3) at (-0.5,-0.75) {};
	\node [targetinode] (v2) at (0.5,-0.75) {};
	
	\draw [neverdiredge, t0] (v1) edge (v2);
	\draw [newdiredge, t1] (v1) edge (v3);
	\draw [olddiredge, t3,] (v2) edge[bend left=20] (v3);
	\draw [olddiredge, t2,] (v3) edge[bend left=20] (v2);
	
	\node [textnode, text=red] at (0,0.409) {-0.164};
\end{scope}

\begin{scope}[local bounding box=bb, shift={(3.5,0.1365)}]  
	\node [sourceinode] (v1) at (0,0) {};
	\node [inode] (v3) at (-0.5,-0.75) {};
	\node [targetinode] (v2) at (0.5,-0.75) {};
	
	\draw [neverdiredge, t0] (v1) edge (v2);
	\draw [olddiredge, t1] (v1) edge[bend left=20]  (v3);
	\draw [newestdiredge, t1] (v3) edge[bend left=20]  (v1);
	\draw [newestdiredge, t1] (v2) edge (v3);
	\node [textnode, text=red] at (0,0.409) {-0.124};
\end{scope}

\begin{scope}[local bounding box=bb, shift={(5.25,0.1365)}]  
	\node [sourceinode] (v1) at (0,0) {};
	\node [inode] (v3) at (-0.5,-0.75) {};
	\node [targetinode] (v2) at (0.5,-0.75) {};
	
	\draw [neverdiredge, t0] (v1) edge (v2);
	\draw [olddiredge, t1] (v1) edge[bend left=20] (v3);
	\draw [newestdiredge, t1] (v3) edge[bend left=20] (v1);
	\draw [olddiredge, t1] (v2) edge[bend left=20] (v3);
	\draw [newestdiredge, t1] (v3) edge[bend left=20] (v2);
	\node [textnode, text=red] at (0,0.409) {-0.115};
\end{scope}

\begin{scope}[local bounding box=bb, shift={(0,-1.909)}]  
	\node [sourceinode] (v1) at (0,0) {};
	\node [inode] (v3) at (-0.5,-0.75) {};
	\node [targetinode] (v2) at (0.5,-0.75) {};
	
	\draw [neverdiredge, t0] (v1) edge (v2);
	\draw [newdiredge, t1] (v1) edge[bend left=20] (v3);
	\draw [newdiredge, t2] (v3) edge[bend left=20] (v1);
	\draw [newdiredge, t2] (v3) edge (v2);
	\node [textnode, text=red] at (0,0.3635) {-0.113};
\end{scope}

\begin{scope}[local bounding box=bb, shift={(1.75,-1.909)}]  
	\node [sourceinode] (v1) at (0,0) {};
	\node [inode] (v3) at (-0.5,-0.75) {};
	\node [targetinode] (v2) at (0.5,-0.75) {};
	
	\draw [neverdiredge, t0] (v1) edge (v2);
	\draw [newestdiredge, t1] (v1) edge[bend left=20] (v3);
	\draw [newestdiredge, t1] (v3) edge[bend left=20] (v1);
	\draw [newestdiredge, t1] (v2) edge (v3);
	
	\node [textnode, text=green] at (0,0.3635) {0.097};
\end{scope}

\begin{scope}[local bounding box=bb, shift={(3.5,-1.909)}]  
	\node [sourceinode] (v1) at (0,0) {};
	\node [inode] (v3) at (-0.5,-0.75) {};
	\node [targetinode] (v2) at (0.5,-0.75) {};
	
	\draw [olddiredge, t1] (v1) edge (v2);
	\draw [olddiredge, t1] (v2) edge[bend left=20] (v3);
	\draw [newdiredge, t1] (v3) edge[bend left=20] (v2);
	\node [textnode, text=red] at (0,0.3635) {-0.088};
\end{scope}

\begin{scope}[local bounding box=bb, shift={(5.25,-1.909)}]  
	\node [sourceinode] (v1) at (0,0) {};
	\node [inode] (v3) at (-0.5,-0.75) {};
	\node [targetinode] (v2) at (0.5,-0.75) {};
	
	\draw [neverdiredge, t0] (v1) edge (v2);
	\draw [newestdiredge, t3] (v1) edge (v3);
	\draw [newestdiredge, t3] (v3) edge (v2);
	\node [textnode, text=green] at (0,0.3635) {0.088};
\end{scope}

\begin{scope}[local bounding box=bb, shift={(0,-3.909)}]  
	\node [sourceinode] (v1) at (0,0) {};
	\node [inode] (v3) at (-0.5,-0.75) {};
	\node [targetinode] (v2) at (0.5,-0.75) {};
	
	\draw [neverdiredge, t0] (v1) edge (v2);
	\draw [olddiredge, t2] (v1) edge[bend left=20] (v3);
	\draw [olddiredge, t2] (v3) edge[bend left=20] (v1);
	\draw [olddiredge, t2] (v2) edge (v3);
	\node [textnode, text=red] at (0,0.3635) {-0.088};
\end{scope}

\begin{scope}[local bounding box=bb, shift={(1.75,-3.909)}]  
	\node [sourceinode] (v1) at (0,0) {};
	\node [inode] (v3) at (-0.5,-0.75) {};
	\node [targetinode] (v2) at (0.5,-0.75) {};
	
	\draw [neverdiredge, t0] (v1) edge (v2);
	\draw [olddiredge, t1] (v1) edge[bend left=20] (v3);
	\draw [olddiredge, t1] (v3) edge[bend left=20] (v1);
	\draw [newestdiredge, t1] (v2) edge [bend left=20] (v3);
	\draw [newdiredge, t1] (v3) edge [bend left=20] (v2);
	\node [textnode, text=red] at (0,0.3635) {-0.087};
\end{scope}

\begin{scope}[local bounding box=bb, shift={(3.5,-3.909)}]  
	\node [sourceinode] (v1) at (0,0) {};
	\node [inode] (v3) at (-0.5,-0.75) {};
	\node [targetinode] (v2) at (0.5,-0.75) {};
	
	\draw [olddiredge, t1] (v1) edge (v2);
	\draw [newestdiredge, t1] (v1) edge[bend left=20] (v3);
	\draw [olddiredge, t1] (v3) edge[bend left=20] (v1);
	\node [textnode, text=red] at (0,0.3635) {-0.081};
\end{scope}

\begin{scope}[local bounding box=bb, shift={(5.25,-3.909)}]  
	\node [sourceinode] (v1) at (0,0) {};
	\node [inode] (v3) at (-0.5,-0.75) {};
	\node [targetinode] (v2) at (0.5,-0.75) {};
	
	\draw [neverdiredge, t0] (v1) edge (v2);
	\draw [olddiredge, t1] (v1) edge[bend left=20] (v3);
	\draw [newdiredge, t1] (v3) edge[bend left=20] (v1);
	\draw [newestdiredge, t1] (v3) edge (v2);
	\node [textnode, text=green] at (0,0.3635) {0.081};
\end{scope}

\begin{scope}[shift={(-1,0.125)}]  
	\node [sourceinode] at (0.775,-5.75) {};
	\node [textnode] at (0.725,-6.175) {\textsf{Source}};
	
	\node [targetinode] at (1.9,-5.75) {};
	\node [textnode] at (1.9,-6.175) {\textsf{Target}};
	
	\node [hidden] (v4) at (2.725,-5.75) {};
	\node [hidden] (v5) at (3.475,-5.75) {};
	\draw [neverdiredge] (v4) edge (v5);
	\node [textnode] at (3.075,-6.175) {\textsf{Never}};
	
	\node [hidden] (v6) at (3.975,-5.75) {};
	\node [hidden] (v7) at (4.725,-5.75) {};
	\draw [newestdiredge] (v6) edge (v7);
	\node [textnode] at (4.3,-6.175) {\textsf{Newest}};
	
	\node [hidden] (v8) at (5.025,-5.75) {};
	\node [hidden] (v9) at (5.775,-5.75) {};
	\draw [newdiredge] (v8) edge (v9);
	\node [textnode] at (5.4,-6.175) {\textsf{New}};
	
	\node [hidden] (v10) at (6,-5.75) {};
	\node [hidden] (v11) at (6.75,-5.75) {};
	\node [textnode] at (6.325,-6.175) {\textsf{Old}};
	\draw [olddiredge] (v10) edge (v11);
	
	\node [hidden] (v10) at (2,-6.5) {};
	\node [hidden] (v11) at (2.75,-6.5) {};
	\node [textnode] at (2.325,-6.925) {\textsf{`0'}};
	\draw [diredge, t0] (v10) edge (v11);
	
	\node [hidden] (v10) at (2.9,-6.5) {};
	\node [hidden] (v11) at (3.65,-6.5) {};
	\node [textnode] at (3.225,-6.925) {\textsf{`1'}};
	\draw [diredge, t1] (v10) edge (v11);
	
	\node [hidden] (v10) at (3.8,-6.5) {};
	\node [hidden] (v11) at (4.55,-6.5) {};
	\node [textnode] at (4.125,-6.925) {\textsf{`2'}};
	\draw [diredge, t2] (v10) edge (v11);
	
	\node [hidden] (v10) at (4.7,-6.5) {};
	\node [hidden] (v11) at (5.45,-6.5) {};
	\node [textnode] at (5.025,-6.925) {\textsf{`3+'}};
	\draw [diredge, t3] (v10) edge (v11);
\end{scope}

\node [textnode] at (0,-1) {\textsf{1}};
\node [textnode] at (1.75,-1) {\textsf{2}};
\node [textnode] at (3.5,-1) {\textsf{3}};
\node [textnode] at (5.25,-1) {\textsf{4}};
\node [textnode] at (0,-3) {\textsf{5}};
\node [textnode] at (1.75,-3) {\textsf{6}};
\node [textnode] at (3.5,-3) {\textsf{7}};
\node [textnode] at (5.25,-3) {\textsf{8}};
\node [textnode] at (0,-5) {\textsf{9}};
\node [textnode] at (1.75,-5) {\textsf{10}};
\node [textnode] at (3.5,-5) {\textsf{11}};
\node [textnode] at (5.25,-5) {\textsf{12}};
\end{tikzpicture}